\documentclass[reprint,superscriptaddress,
nofootinbib,
 amsmath,amssymb,
 aps,
pra,
]{revtex4-2}

\usepackage[utf8]{inputenc} 
\usepackage[T1]{fontenc}    
\usepackage{hyperref}       
\usepackage{url}            
\usepackage{booktabs}       
\usepackage{tabularx}
\usepackage{multirow}
\usepackage{amsfonts}       
\usepackage{nicefrac}       
\usepackage{microtype}      
\usepackage{graphicx}
\usepackage{xcolor, etoolbox, colortbl}
\usepackage{amsmath,amsfonts,amsthm} 
\usepackage{mathtools}
\usepackage{ragged2e}
\usepackage{braket}      
\usepackage[english]{babel} 
\graphicspath{{./figures/}} 
\usepackage{enumitem}
\usepackage{xspace}
\usepackage{dsfont}
\usepackage{bm}
\usepackage{bbm}
\usepackage{dcolumn}
\usepackage{mathrsfs}

\newcommand{\Tr}{\textbf{Tr}}
\newcommand{\tr}{\text{Tr}}

\definecolor{Gray}{gray}{0.95}

\hypersetup{
    colorlinks,
    linkcolor=[rgb]{0.206756, 0.371758, 0.553117},
    urlcolor=[rgb]{0.120565, 0.596422, 0.543611},
    citecolor=[rgb]{0.8784313725490196, 0.403921568627451, 0.4549019607843137},
    breaklinks=true 
}

\DeclarePairedDelimiter\abs{\lvert}{\rvert}

\DeclarePairedDelimiter\floor{\lfloor}{\rfloor}

\begin{document}

\title{Configurable photonic simulator for quantum field dynamics}

\author{Mauro D'Achille}
    \email{mauro.dachille@uni-jena.de}
    \affiliation{Institute of Condensed Matter Theory and Optics, Friedrich-Schiller-University Jena, Max-Wien-Platz 1, 07743 Jena, Germany}

\author{Martin G\"{a}rttner}
    \email{martin.gaerttner@uni-jena.de}
    \affiliation{Institute of Condensed Matter Theory and Optics, Friedrich-Schiller-University Jena, Max-Wien-Platz 1, 07743 Jena, Germany}

\author{Tobias Haas}
    \email{tobias.haas@uni-ulm.de}
    \affiliation{Institut für Theoretische Physik and IQST, Universität Ulm, Albert-Einstein-Allee 11, 89069 Ulm, Germany}
    \affiliation{Centre for Quantum Information and Communication, École polytechnique de Bruxelles, CP 165, Université libre de Bruxelles, 1050 Brussels, Belgium}

\begin{abstract}
Quantum field simulators provide unique opportunities for investigating the dynamics of quantum fields through tabletop experiments. A primary drawback of standard encoding schemes is their rigidity: altering the theory, its coupling geometry, metric structure, or simulation time typically requires redesigning the experimental setup, which imposes strong constraints on the types of dynamics and theories that can be simulated. Here, we introduce the Optical Time Algorithm (OTA) as a unifying framework, enabling the efficient simulation of large classes of free quantum field dynamics using a single optical circuit design that separates the time from the Hamiltonian's structure. By modifying the parameters of the optical elements, our method allows us to engineer timescales, coupling graphs, spacetime metrics, and boundary conditions, thereby facilitating the implementation of relativistic and non-relativistic, real- and complex-valued, short- and long-range quantum field theories on both flat and curved spacetimes. We exploit the OTA's configurability to investigate the spreading of quantum correlations in space and time for theories with continuously varying coupling ranges. Relevant features predicted by quantum field theory can be observed in systems with $10$ to $20$ modes under realistic conditions, paving the way for experimental implementations.
\end{abstract}

\maketitle

\section{Introduction}
Over 40 years after Unruh's proposal for mimicking Hawking radiation with sound waves~\cite{Unruh1981} and Feynman's insight that quantum physics is best simulated by quantum computers~\cite{Feynman1982}, quantum field simulators have become a central tool for probing the non-equilibrium dynamics of quantum field theories~\cite{Georgescu2014,Altman2021,Bauer2023,Halimeh2025}. By virtue of their high-level of control in the quantum regime, much attention has been drawn to ultracold atoms~\cite{Bloch2012,Periwal2021,Haas2022b,Tajik2023b,Haas2025b},
trapped-ions~\cite{Blatt2012,Brydges2019,Monroe2021}, and Rydberg systems~\cite{Browaeys2020,Bluvstein2021,GonzalezCuadra2025}, all of which allow for probing few- and many-body physics. While each platform offers distinct opportunities, common approaches share three main limitations: they operate far from the field-theory limit, are often constrained by fixed-coupling geometries, and allow coherent dynamics only on short time scales. This poses significant challenges for simulating the dynamics of entire classes of proper field theories within a \textit{single} setup.

Also photonic systems have long been considered for quantum simulation~\cite{AspuruGuzik2012,Noh2016,Somhorst2023} and computing~\cite{Knill2001,Kok2007,Romero2025} tasks. In recent years, the community has put a strong focus on quantum-advantage tests based on boson sampling, using either Fock-state inputs~\cite{Aaronson2013} or Gaussian inputs~\cite{Hamilton2017,Kruse2019}. Remarkably, Gaussian boson sampling operates entirely in non-interacting, i.e., Gaussian, regimes until photon-number detection at the output---yet, sampling from the outcome distribution is widely believed to pose a $\#$P-hard problem in general~\cite{Hangleiter2023}. The strongest experimental indications for quantum advantage have so far emerged for the Gaussian variant, with Jiuzhang and successive upgrades~\cite{Zhong2020,Zhong2021,Deng2023,Liu2025} as well as Borealis~\cite{Madsen2022}. These demonstrations highlight what current photonic hardware excels at: large mode numbers and Hilbert-space dimensions ($\sim 10^2$–$10^4$ modes~\cite{Liu2025}), highly programmable Gaussian unitaries (reaching $\sim95 \%$ similarity between measured and target amplitude distributions~\cite{Stefszky2025}), and fast sampling rates (up to $\sim10^6$ samples/s~\cite{Stefszky2025}). At the same time, they emphasize photonics' main bottleneck: loss is ubiquitous, and sufficiently lossy devices can bring medium-sized instances back within reach of efficient classical algorithms~\cite{Oh2024a,Oh2024b}.

Given the unique capabilities of photonic platforms, especially when interferometers are integrated into programmable devices~\cite{Pelucchi2022,Wang2025,AghaeeRad2025,PsiQuantum2025}, simulating quantum fields with light has emerged as a compelling alternative to current approaches. In contrast to simulators based on spin degrees of freedom, photons with their bosonic nature are better suited for implementing scalar field theories, thus overcoming the need for digitization~\cite{Jordan2012,Jordan2014,Klco2019}. Further, leveraging the programmability of photonic circuits enables simulating quantum fields in curved spacetimes beyond the constraints of Unruh's analog-gravity paradigm~\cite{Barcelo2011,Braunstein2023}, which relies on fluid-like analogies in implementations with ultracold atoms~\cite{Eckel2018,Haas2022b,Steinhauer2014,MunozDeNova2019,Hu2019}, optical fibers~\cite{Philbin2008}, or quantum fluids~\cite{Steinhauer2021,Svancara2024}.

While early works have proposed optical circuits for ground-state preparation and scattering~\cite{Marshall2015} as well as for the dynamics of specific theories~\cite{Kalajdzievski2018,Abel2024}, the missing piece of the puzzle is a versatile yet efficient algorithm that maps large classes of quantum field dynamics onto optical components, taking into account current and near-term experimental capabilities. Considering free theories characterized by quadratic Hamiltonians, many black-box decompositions into optical elements are already known, see~\cite{Houde2024} for a recent overview. The arguably most prominent variant is the Bloch-Messiah decomposition~\cite{Bloch1962}, which consists of a single-mode squeezer layer augmented with two all-to-all interferometers (the latter further decomposes into many phase shifts and beam splitters~\cite{Reck1994,Clements2016}). However, this and other decompositions standard in quantum optics come with a major drawback hindering their prospects for quantum field simulation: \textit{all} components, for instance, all squeezers and interferometers in the case of Bloch-Messiah, must be entirely reconfigured not only depending on the theory of interest, but also for each time step of a quantum dynamics simulation. This poses severe challenges for free-space optics, where reconfiguration typically requires repositioning and restabilizing individual components, but also critically slows down the sampling speed of integrated photonic processors---from the $\mu$s-regime~\cite{Stefszky2025} to $1-10$s~\cite{Taballione2023,Fyrillas2024}, corresponding to a $10^6-10^7$-fold increase.

In this work, we propose the Optical Time Algorithm (OTA) as a unified framework for simulating free scalar field theories on photonic platforms. In contrast to existing methods, our decomposition into optical elements has a clear and physically intuitive structure: while a fixed interferometer array followed by a single layer of squeezers implements a largely arbitrary coupling geometry, the dynamics is \textit{isolated} in a single layer of phase shifters. This facilitates the simulation of the non-equilibrium dynamics of various quantum field theories without the need to ever alter the experimental design.

Remarkably, we find that, for initial product coherent states, the OTA reduces to the Gaussian boson sampler setup. Therefore, our algorithm not only opens programmable quantum field simulation as a novel use case for state-of-the-art sampling experiments, but also provides a pathway to practical quantum advantage, given the classical hardness of assessing the full particle-number distribution.

As an application of our method, we investigate information dynamics in terms of entanglement entropy and quantum mutual information~\cite{Plenio2007,Adesso2012}, also for non-local theories. We show that analytic predictions in the continuum for how quantum entanglement~\cite{Amico2008,Eisert2010} spreads in spacetime based on the quasi-particle picture~\cite{Calabrese2005,Calabrese2020,Modak2020} are in agreement with simulations with $10-20$ modes and remain robust to photon loss and noise. 

\textit{The remainder of this paper is structured as follows}. We set the stage by recalling the necessary tools for describing continuous-variable quantum systems in Sec.~\ref{sec:Preliminaries}. We present our main result, the OTA [see Eq.~\eqref{eq:OTA}], in Sec.~\ref{sec:Decomposition}, and discuss its relation to Gaussian boson sampling. In Sec.~\ref{sec:Simulation}, we showcase the OTA's versatility by detailing how five rather different types of field theories can be simulated by minimally adjusting circuit parameters. As an application, we investigate the dynamics of quantum correlations after a quench in experimentally accessible regimes in Sec.~\ref{sec:Applications}, followed by an extensive analysis of experimental imperfections in Sec.~\ref{sec:Error analysis}. We discuss our findings in Sec.~\ref{sec:Discussion}.

\textit{Notation}. We use natural units $\hbar = c = k_{\text{B}} = G = 1$, denote quantum operators (classical variables) by bold (normal) letters $\boldsymbol{\rho}$ ($\gamma$), analogously for their traces $\Tr \{ \boldsymbol{\rho} \}$ ($\tr \{ \gamma \}$) and work in $(1,d)$-dimensional spacetime with metric signature $(+,-, \dots, -)$.

\section{Preliminaries}
\label{sec:Preliminaries}
We begin by mapping quantum fields to optical modes and introducing fundamental concepts in quantum optics~\cite {Weedbrook2012,Serafini2017}. In particular, we introduce Gaussian operations and optical elements---phase shifters, beam splitters, and squeezers---serving as fundamental building blocks.

\subsection{From quantum fields to discrete modes}
\begin{figure}[t!]
    \centering
    \includegraphics[clip,width=1.\columnwidth]{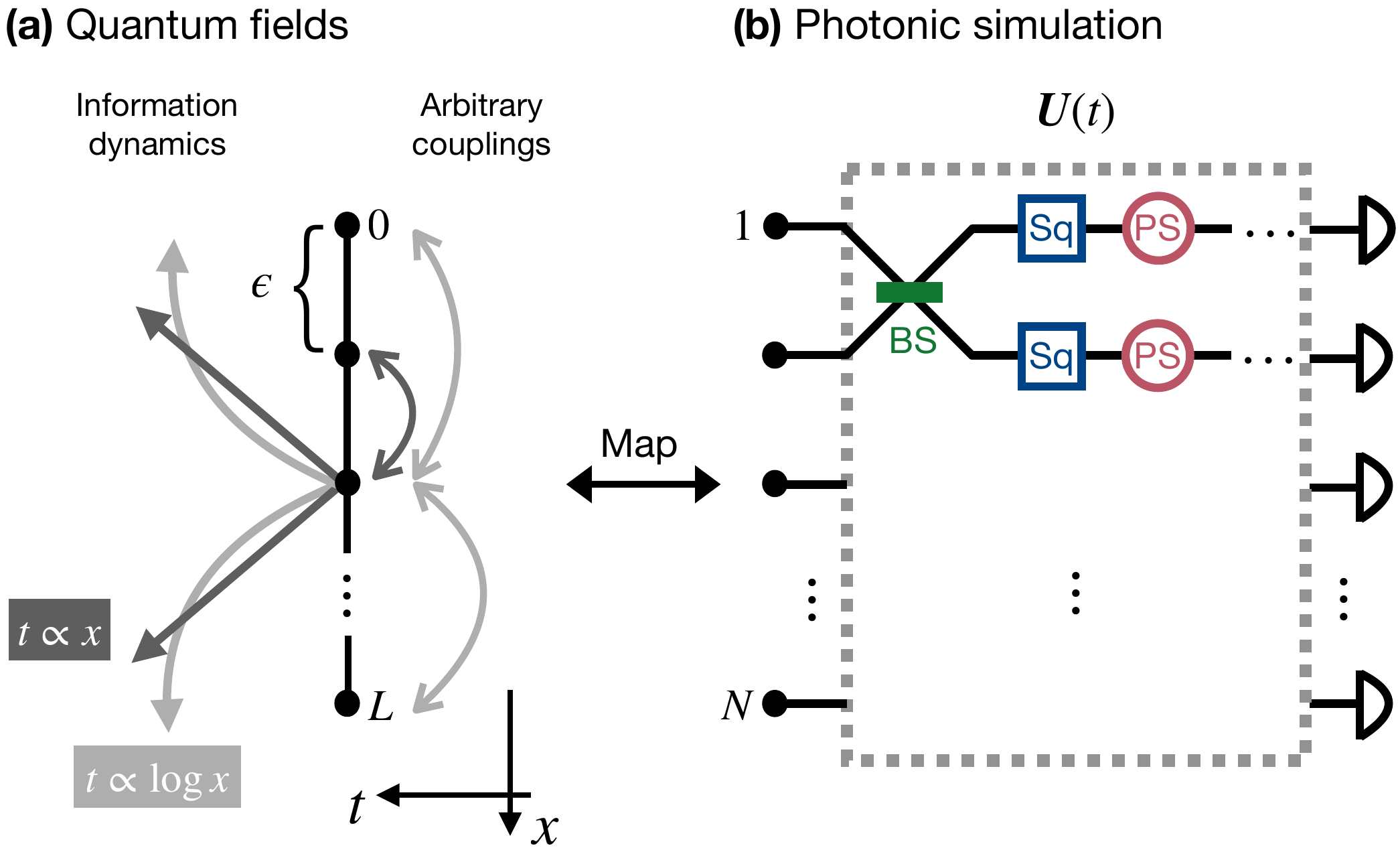}
    \caption{Pictorial representation of the map between scalar quantum fields \textbf{(a)} and their photonic simulation \textbf{(b)}. By tuning the optical parameters, one can realize arbitrary coupling geometries and timescales, thereby enabling the probing of information dynamics in spacetime in general settings.}
    \label{fig:Overview}
\end{figure}

We consider a real scalar quantum field $\phi (x)$ together with its conjugate momentum field $\pi (x)$ over a finite hypercube $x \in [0, L]^d$. We discretize spatial positions by introducing a hypercubic lattice with spacing $\epsilon > 0$ and label discrete spatial modes with $j \in \mathcal{J} = \{1, \ldots, \tilde{N}\}^d$, such that $L = \tilde{N} \epsilon$, see Fig.~\hyperref[fig:Overview]{1(a)}. We choose both induced discretized fields to be dimensionless, i.e.,
\begin{equation} \label{eq:Rescaled Fields in Position Space}
    \phi(x) \rightarrow \phi_{j} = \phi(\epsilon j), \quad \pi(x) \rightarrow \pi_{j} = \epsilon^d \, \pi(\epsilon j),
\end{equation}
by means of a non-symmetric rescaling. This ensures that the corresponding quantum operators fulfill the standard canonical commutation relations (CCRs)
\begin{equation}
    [\boldsymbol{\phi}_{j} , \boldsymbol{\pi}_{k}] = i \delta_{j k} \mathds{1},
    \label{eq:CCRStandard}
\end{equation}
and agrees with quantum optics conventions. For simplicity, we relabel spatial indices to run from $1$ to $N$, where $N=\tilde{N}^d$ denotes the total number of modes.

We group\footnote{Note that another prominent convention uses the grouping $\boldsymbol{r} = (\boldsymbol{\phi}_1, \boldsymbol{\pi}_1, \dots, \boldsymbol{\phi}_N, \boldsymbol{\pi}_N)^{\intercal}$ \cite{Weedbrook2012,Serafini2017}.} the fields into a single vector $\boldsymbol{r} = (\boldsymbol{\phi}_1, \dots, \boldsymbol{\phi}_N, \boldsymbol{\pi}_1, \dots, \boldsymbol{\pi}_N)^{\intercal}$. Then, the CCRs~\eqref{eq:CCRStandard} become
\begin{equation}
    [ \boldsymbol{r} , \boldsymbol{r}^{\intercal} ] = i \Omega,
    \label{eq:CCRCompact}
\end{equation}
where $\Omega$ is the so-called symplectic form,
\begin{equation}
    \Omega =
    \begin{pmatrix}
        \mathds{O}_{N} & \mathds{1}_{N} \\
        - \mathds{1}_{N} & \mathds{O}_{N}
    \end{pmatrix}
    = (i \sigma_{y}) \otimes \mathds{1}_{N} ,
\end{equation}
with $\sigma_{y}$ being the second Pauli matrix and $\mathds{O}_{N}$ ($\mathds{1}_{N}$) the $N \times N$ null (identity) matrix. The symplectic form is orthogonal and involutory up to a sign, $\Omega^{-1} = \Omega^{\intercal} = - \Omega$.

\subsection{Gaussian unitaries and symplectic matrices}
Free, i.e., non-interacting, field theories are characterized by second-order Hamiltonians, which result in Gaussian unitaries, as they map Gaussian states---states with a Gaussian Wigner function---to Gaussian states. In this case, the most general Hamiltonian takes the form
\begin{equation}
    \boldsymbol{H} = \dfrac{1}{2} \boldsymbol{r}^{\intercal} H \boldsymbol{r} + \boldsymbol{r}^{\intercal} s,
\end{equation}
where $H$ denotes the symmetric and positive semi-definite Hamiltonian matrix and $s$ is a real $2N$-dimensional vector. Without loss of generality, we consider purely quadratic Hamiltonians and set $s \equiv 0$, since a linear term can be accounted for by a simple displacement.

The corresponding unitary time-evolution operator is
\begin{equation}
    \boldsymbol{U}(t) = e^{-it \boldsymbol{H}} = e^{-\frac{i t}{2} \boldsymbol{r}^{\intercal} H \boldsymbol{r}}.
    \label{eq:TimeEvolutionUnitary}
\end{equation}
The action of this unitary onto field operators can be equally described by symplectic matrices via $\boldsymbol{U}^{\dagger} \boldsymbol{r} \boldsymbol{U} = S \boldsymbol{r}$, where we introduced the symplectic transformation 
\begin{equation} \label{Definition of symplectic matrix}
    S (t) = e^{\Omega H t}.
\end{equation}
In general, a matrix $S$ is called symplectic if it preserves the CCRs~\eqref{eq:CCRCompact}, i.e., $S \Omega S^{\intercal} = \Omega$, and the set of symplectic matrices forms the symplectic group $Sp(2N, \mathbb{R})$. Note also that they have unit determinant $\det S = 1$.

A fundamental result for our analysis is the normal-mode decomposition, also known as Williamson's theorem~\cite{Williamson1935}. Given a $2N \times 2N$ real and positive semi-definite matrix $H$ (e.g., the Hamiltonian matrix), there exists a symplectic diagonalizing (SD) transformation $S_{\text{SD}} \in Sp(2N, \mathbb{R})$ in the sense that~\cite{Son2022, Nicacio2021}
\begin{equation}
    S_{\text{SD}}^{\intercal} D S_{\text{SD}} = H
    \, \, \,  \text{where} \, \, \, 
    D = D_{N} \oplus D_{N} ,
    \label{eq:WilliamsonsTheorem}
\end{equation}
with the diagonal matrix
\begin{equation}
    D_{N} = \text{diag}(d_{1}, \ldots, d_{N}) ,
\end{equation}
containing the so-called symplectic eigenvalues $d_{j} \in \mathbb{R}$. Importantly, the SD matrix is not unique~\cite{deGosson2006,Son2021}: the product of two SD matrices satisfies $S_{\text{SD}} S_{\text{SD}}'^{-1} = G \in OrSp(2N)$, where
\begin{equation}
    \begin{split}
        OrSp(2N) & = \Big\{
        G = \begin{pmatrix}
            G_{1} & G_{2} \\
            -G_{2} & G_{1}
        \end{pmatrix}: 
        G_{1}^{\intercal} G_{2} = G_{2}^{\intercal} G_{1}, \\
        & \hspace{0.7cm}
        G_{1}^{\intercal} G_{1} + G_{2}^{\intercal} G_{2} = \mathds{1}_{N}
        \Big\} 
    \end{split}
\end{equation}
defines the orthosymplectic group. It includes all transformations that are simultaneously orthogonal and symplectic, i.e., passive transformations.

\subsection{Optical elements}
Finally, we introduce the fundamental optical elements that are relevant for a photonic simulation, see Fig.~\hyperref[fig:Overview]{1(b)}, which are conveniently grouped into passive (energy-conserving) and active (energy-changing) operations: 
\begin{enumerate} [label = (\roman*)]
    \item The (passive) single-mode phase shift (PS)
    \begin{equation}
        H_{\text{PS}} = \mathds{1}_{2}, \quad S_{\text{PS}}(\varphi)
        = \begin{pmatrix}
            \cos{\varphi} & \sin{\varphi} \\
            -\sin{\varphi} & \cos{\varphi}
        \end{pmatrix},
        \label{eq:PhaseShift}
    \end{equation}
    which implements a phase rotation $\varphi \in [0, 2\pi).$
    
    \item The (passive) two-mode beam splitter (BS)
    \begin{equation}
        \begin{split}
            H_{\text{BS}} &= \sigma_{y} \otimes \sigma_{y}, \\
            S_{\text{\text{BS}}}(\theta) &= \begin{pmatrix}
            \cos{\theta} & \sin{\theta} & 0 & 0 \\
            -\sin{\theta} & \cos{\theta} & 0 & 0 \\
            0 & 0 & \cos{\theta} & \sin{\theta}\\
            0 & 0 & -\sin{\theta} & \cos{\theta}
        \end{pmatrix},
        \label{S transformation for BS}
        \end{split}
    \end{equation}
    describing the interference of two modes with transmissivity $\cos^{2}(\theta)$ controlled by $\theta \in [-\pi,\pi]$\footnote{A physical beam splitter is described by $\theta \in [0,\pi/2]$. The remaining values can be realized with two additional $\pi$ phase shifts: one on the first mode before the BS, and one after the BS on mode two for $\theta \in (\pi/2, \pi]$, one before on mode two for $\theta \in [-\pi,-\pi/2)$, and one after on mode one when $\theta \in [-\pi/2,0)$.}.
    \item The (active) single-mode squeezer (Sq)
    \begin{equation}
        H_{\text{Sq}} = - \sigma_{x}, \quad S_{\text{Sq}}(z) 
        = \begin{pmatrix}
            e^{-z} & 0 \\
            0 & e^{z}
        \end{pmatrix},
        \label{eq:Squeezer}
    \end{equation}
    which decreases the uncertainty in one of the two fundamental fields at the cost of equally increasing the other, governed by $z \in \mathbb{R}$.
\end{enumerate}
For multi-mode systems, we denote by $S_{\text{PS}}(\varphi)$ the $N$-mode phase shift with phase vector $\varphi = (\varphi_1, \dots, \varphi_N)^{\intercal}$ composed out of single-mode phase shifts $\varphi_j$; analogously for the $N$-mode squeezer.

\section{Optical time algorithm}
\label{sec:Decomposition}
We are now ready to state and discuss our main result, the Optical Time Algorithm (OTA).

\subsection{Main result}

\subsubsection{Isolating time dependencies}
Building upon Williamson's theorem, cf.\ Eq.~\eqref{eq:WilliamsonsTheorem}, the symplectic time evolution matrix $S(t)$ corresponding to some Hamiltonian matrix $H$ can always be decomposed as (see App.~\ref{subsec:Proof1} for a proof)
\begin{equation} 
    S(t) = S_{\text{SD}}^{-1} \, S_{\text{PS}}(\varphi(t)) \, S_{\text{SD}} .
    \label{eq:IsolatingTime}
\end{equation}
Here, the phases $\varphi(t) = t \, (d_{1}, \ldots, d_{N})^{\intercal}$ carry all time-dependencies in a linear fashion, and $d_j$ denote the symplectic eigenvalues of $H$. Thus, the dynamics is entirely encoded in a single layer of phase shifters. 

\subsubsection{Decomposition into optical elements}
\label{subsubsec:OTA2}
\begin{figure}[t!]
\centering
    \includegraphics[clip,width=0.85\columnwidth]{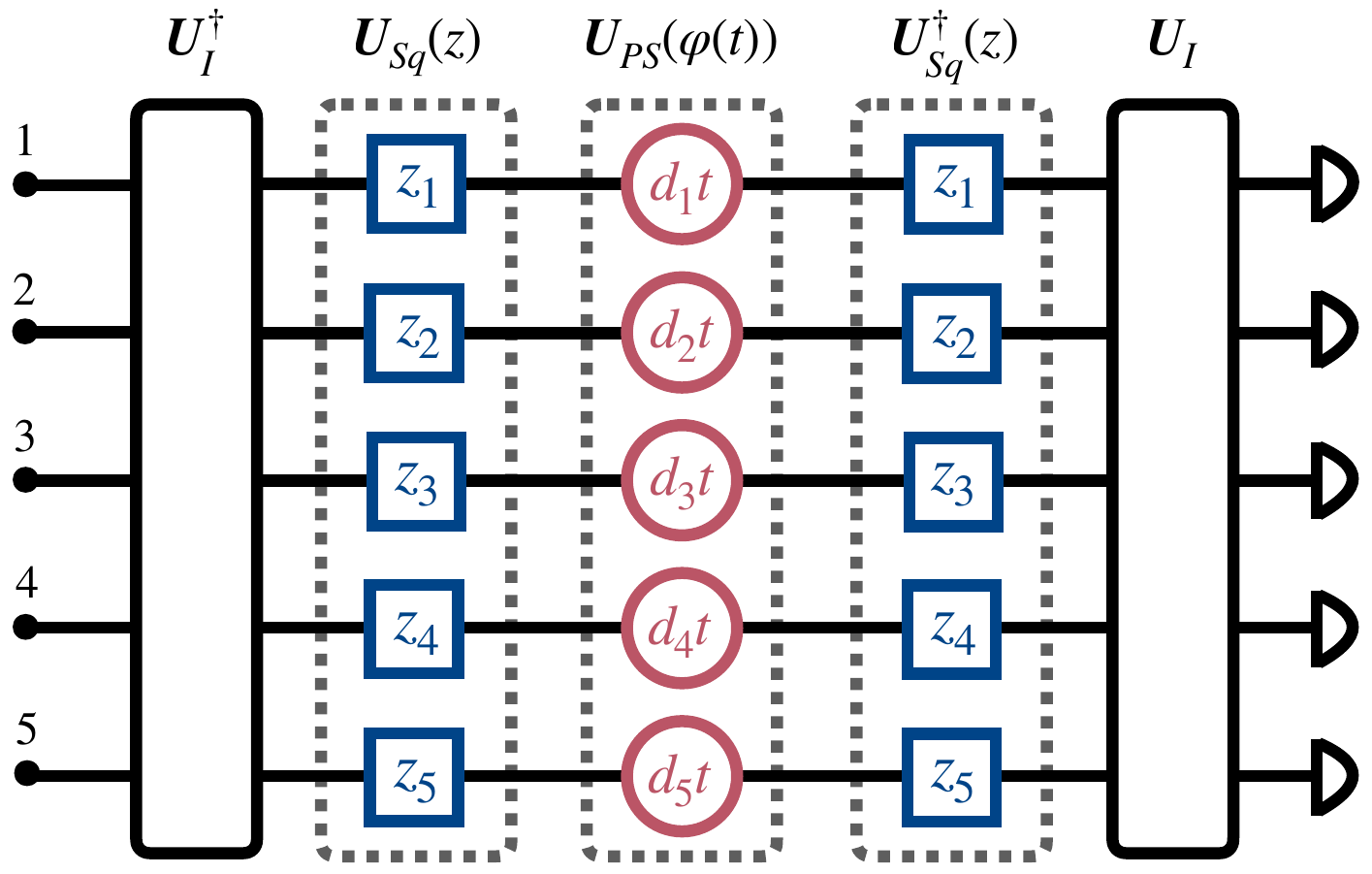}
    \caption{Pictorial representation of the OTA for $N=5$. The first and last layers can be further decomposed via standard techniques, such as Reck's~\cite{Reck1994} and Clements' decompositions~\cite{ Clements2016}. We provide an explicit example for the former in App.~\ref{subsubsec:BeamSplitterArray}.} 
    \label{fig:OTA}
\end{figure}

For the vast majority of physically relevant scalar quantum field theories, the Hamiltonian matrix is of block-diagonal form $H = H^{\phi} \oplus H^{\pi}$ with commuting blocks $[H^{\phi}, H^{\pi}] = \mathds{O}_{N}$. Under this assumption, we prove in App.~\ref{subsec:Proof2} that $S(t)$ admits the general decomposition
\begin{equation}
    S(t) = S_{\text{I}} \, S_{\text{Sq}}^{-1}(z) \, S_{\text{PS}}(\varphi(t)) \, S_{\text{Sq}}(z) \, S_{\text{I}}^{-1} ,
    \label{eq:OTA}
\end{equation}
where $S_{\text{I}}$ is a time-independent passive transformation, here denoted as interferometer.
Remarkably, the design of the corresponding optical circuit, see Fig.~\hyperref[fig:OTA]{2}, is independent of the theory under consideration. Instead, the structure of the Hamiltonian of interest is fully encoded in the parameters of the optical elements via
\begin{subequations}
    \begin{align}
        & \varphi(t) = t \, (d_{1}, \ldots, d_{N})^{\intercal} , \label{eq:OTAParameters1} \\
        & z = - (\ln\gamma_{1}, \ldots, \ln \gamma_{N})^{\intercal}/2 \label{eq:OTAParameters2}, \\
        & S_{\text{I}}^{-1}
        = P_{N} \oplus P_{N} \label{eq:OTAParameters3}.
    \end{align}
\end{subequations}
Therein, $P_{N} = [ p_{1}, \ldots, p_{N} ]^{\intercal}$ is composed of the real and orthonormal eigenvectors $p_{j}$ of $H^{\phi}$ (and $H^{\pi})$. The ratio of the corresponding eigenvalues $\lambda^{\phi}_{j}$ and $\lambda^{\pi}_{j}$ is related to the squeezing amplitudes
\begin{equation} \label{eq:Eigenvalues ratio}
    \gamma_{j}^{2} = \lambda^{\phi}_{j} / \lambda^{\pi}_{j},
\end{equation}
while the phase shifts and symplectic eigenvalues $d_j$ of $H$ are given by
\begin{equation} \label{eq:Eigenvalues product}
    d_{j}^{2} = \lambda^{\phi}_{j} \lambda^{\pi}_{j}.
\end{equation}
In App.~\ref{subsubsec:BeamSplitterArray}, we consider one of the possible schemes to further decompose the interferometer layers with beam splitters only.
There, we also show how gauge redundancies can be used to minimize the number of gates.

Given a Hamiltonian matrix $H$, all circuit parameters can be calculated using a dedicated software package we made publicly available under~\cite{GitHub}.

The intuition behind the OTA is straightforward: the first interferometer layer diagonalizes the Hamiltonian matrix, thereby decoupling the modes. The following squeezer layer equalizes the prefactors of the fields $\phi$ and $\pi$, after which the (now independent) modes are evolved in time by phase shifters. Then, the second pair of a squeezer and interferometer layer transforms back to the original basis. As we will discuss in Sec.~\ref{sec:Simulation}, the phase shift parameters $d_j$ often reduce to the dispersion relation $\omega_j$ of the field theory of interest. 

Further, we remark that the number of gates required by the OTA scales quadratically with the number of modes $N$, i.e.,
\begin{equation}
    n_{\text{G}}
    = n_{\text{Sq}} + n_{\text{PS}} + n_{\text{I}}
    \sim N (N + 2),
\end{equation}
consistent with other unitary decompositions in terms of single- and two-mode gates~\cite{Nielsen2010}, since each interferometer layer requires $\sim N(N-1)/2$ gates for the decomposition. However, in strong contrast to other decompositions, only the phase shifts have to be tuned to simulate dynamical evolution, while all other components remain time-independent.

\subsubsection{Time-dependent Hamiltonians}
\label{subsubsec:Time-dependent OTA}
When the Hamiltonian carries explicit time dependence $\boldsymbol{H}=\boldsymbol{H}(t)$, the time evolution operator in Eq.~\eqref{eq:TimeEvolutionUnitary} becomes
\begin{equation}
    \boldsymbol{U}(t)
    = \mathcal{T}\exp 
    \bigg[ -i \int_{0}^{t} \mathrm{d}t'
    \boldsymbol{H}(t') \bigg],
\end{equation}
where $\mathcal{T}$ denotes the time-ordering operator. In general, the latter prevents isolating the time in a single phase-shift layer as in Eq.~\eqref{eq:IsolatingTime} since the symplectic diagonalization becomes time-dependent as well $S_{\text{SD}} = S_{\text{SD}}(t)$.

Remarkably, there are relevant cases in which the general form of the OTA, Eq.~\eqref{eq:OTA}, remains intact. When $[\boldsymbol{H}(t_{1}),\boldsymbol{H}(t_{2})]=0 \,\, \forall t_{1}, t_{2}$, which amounts to time-independent eigenvectors $P \neq P(t)$, time-ordering simplifies to the identity $\mathcal{T}=\mathds{1}$. This applies, for instance, to circulant Hamiltonian matrices, which are diagonalizable via a time-independent discrete Fourier transformation,
see Sec.~\ref{subsec:HamiltonianStructure}. Consequently, the interferometer remains time-independent $S_{\text{I}} \neq S_{\text{I}}(t)$, while the phases and squeezing parameters generalize to
\begin{subequations}
    \begin{align}
        & \varphi(t) = (\tilde{d}_{1} (t), \ldots, \tilde{d}_{N} (t))^{\intercal} , \label{eq:Time-dependent OTAParameters1} \\
        & z(t) = - (\ln\tilde \gamma_{1} (t), \ldots, \ln \tilde \gamma_{N} (t))^{\intercal}/2 \label{eq:Time-dependent OTAParameters2},
    \end{align}
\end{subequations}
respectively, see App.~\ref{subsubsec:Proof for time-dependent OTA} for a proof. Here, $\tilde{d}_j$ and $\tilde{z}_j$ encode the now time-dependent eigenvalues 
\begin{equation}
    \tilde{\lambda}_{j}^{\phi(\pi)}(t)
    = \int_{0}^{t} \mathrm{d}t' \lambda_{j}^{\phi(\pi)}(t') .
\end{equation}
via Eqs.~\eqref{eq:Eigenvalues ratio} and Eqs.~\eqref{eq:Eigenvalues product}, respectively. Importantly, only the single-mode optical elements carry time dependence.

\subsection{Connection to Gaussian boson sampling}
\label{subsec:GaussianBosonSampling}

\subsubsection{OTA with a single squeezer layer}
\label{subsubsec:OTA with one squeezer layer}
The obtained decomposition can be modified using known decompositions and adapted to specific experimental constraints. For instance, in order to reduce the number of squeezers, one can apply the Bloch-Messiah decomposition (see Ref.~\cite{Houde2024} for details) to the single-mode part of the OTA, leading to
\begin{equation}
    S(t)
    = S_{\text{I}}
    S_{\text{PS,1}} (t)
    S_{\text{Sq}} (t)
    S_{\text{PS,2}} (t)
    S_{\text{I}}^{-1} .
\end{equation}
While the interferometer layers remain time-independent, both the squeezer and phase shifter arrays acquire time dependencies. This differs from the result obtained by directly applying the Bloch-Messiah decomposition, in which case all gates are generally time-dependent. In fact, well-established Bloch-Messiah decomposition algorithms do not rely on standard or symplectic diagonalization to isolate the time parameter~\cite{Houde2024}. 

\subsubsection{Vacuum inputs}
\label{subsubsec:Simplification for vacuum inputs}
So far, we have worked at the level of time-evolution operators without assuming a specific input state. Interestingly, the OTA requires fewer gate layers when considering the default input state in quantum-optics experiments, namely the product-form vacuum $\ket{0}$ annihilated by all $\boldsymbol{a}_{j} = (\boldsymbol{\phi}_j + i \boldsymbol{\pi}_j)/\sqrt{2}$ (see App.~\ref{subsec:CSq OTA Appendix} for generalizations to coherent inputs). Noting that passive transformations map vacuum to vacuum~\cite{Weedbrook2012}, the first interferometer layer reduces to the identity. Further, for vacuum inputs, the single-mode part can be summarized as complex squeezers, defined by
\begin{equation}
    \begin{split}
    & H_{\text{CSq}}(\Phi)
    = - \sigma_{x} \cos \Phi 
    - \sigma_{z} \sin \Phi , \\
    & S_{\text{CSq}}(\zeta)
    = \cosh \xi \, \mathds{1}_{2}
    - \sinh \xi 
    \begin{pmatrix}
        \cos \Phi & \sin \Phi \\
        \sin \Phi & - \cos \Phi 
    \end{pmatrix} ,
    \end{split}
    \label{eq:ComplexSqueezer}
\end{equation}
where $\zeta = \xi e^{i\Phi}$ encodes amplitude $\xi \in \mathbb{R}$ and phase $\Phi \in [0,\pi)$. The latter reduces to a real squeezer, see Eq.~\eqref{eq:Squeezer}, when $\Phi = 0$, in which case $\zeta = \xi = z$. The optical parameters in Eqs.~\eqref{eq:OTAParameters1} and~\eqref{eq:OTAParameters2} are related with the ones of the complex squeezers via
\begin{equation}\label{eq:Complex Squeezer Parameters}
    \begin{split}
        \cosh(2\xi) &= \cos^{2} (\varphi) + \sin^{2} (\varphi) \cosh(4 z), \\
        \tan \Phi &= \dfrac{\cot \varphi}{\cosh(2 z)},
    \end{split}
\end{equation}
see App.~\ref{subsec:CSq OTA Appendix} for details. The price to pay is that the complex squeezer parameters become time-dependent, i.e., $\xi = \xi (t), \Phi = \Phi (t)$, since they all depend on the phases $\varphi (t)$. When considering time-dependent Hamiltonians of the form discussed in Sec.~\ref{subsubsec:Time-dependent OTA}, this structure is preserved without any additional time dependencies. Remarkably, the structure of the circuit also extends to general time-dependent Hamiltonian matrices $H = H (t)$, in which case the interferometer encodes the entire dynamical history via the time-ordering operator $\mathcal{T}$.

\begin{figure}[t!]
 \centering
    \includegraphics[clip,width=.85\columnwidth]{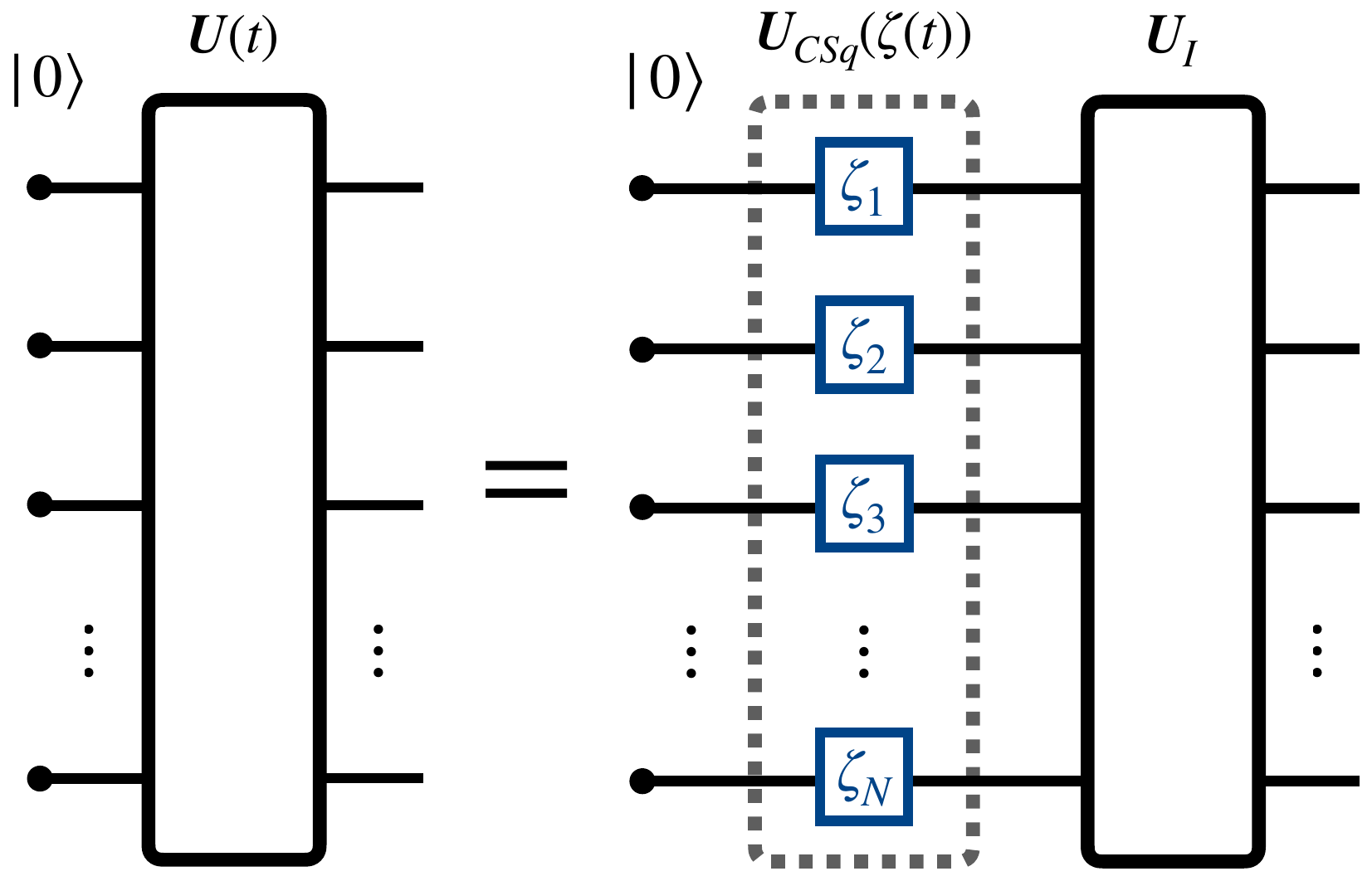}
    \caption{For vacuum inputs, the OTA reduces to a complex squeezer layer followed by an interferometer. The complex-valued squeezing parameters $\zeta = \xi e^{i \Phi}$ are time-dependent, while the interferometer remains time-independent. This matches the structure of the Gaussian boson sampler setup.}
    \label{fig:Boson sampler}
\end{figure}

Thanks to these simplifications, the OTA reduces to a layer of time-dependent squeezers and a \textit{single} time-independent interferometer array, to wit
\begin{equation} \label{eq:Boson sampler simplification}
    S(t) = S_{\text{I}}
    S_{\text{CSq}}(\zeta(t)).
\end{equation}
We sketch the corresponding circuit in Fig.~\ref{fig:Boson sampler}, which precisely matches the one realized in Gaussian boson samplers~\cite{Hamilton2017,Kruse2019}. In such experiments, particle-number-resolving detectors at the output~\cite{Marsili2013,Reddy2020,Cheng2023} introduce non-Gaussianity. It is well-known that the corresponding photon number output distribution $p (n_1, \dots, n_N) = \abs{\braket{n | \boldsymbol{U}_{\text{I}} \boldsymbol{U}_{\text{CSq}} (t) | 0}}^2$ contains a Hafnian~\cite{Hamilton2017}. Computing Hafnians (and even approximating them) is $\# \text{P}$-hard in general, with the computational complexity scaling as $\mathcal{O}(n^3 2^n)$ in total detected photon number $n$~\cite{Bjorklund2019,Hangleiter2023}. Hence, such quantum-optics experiments offer a true quantum advantage over classical algorithms, as recently demonstrated with high confidence~\cite{Madsen2022,Liu2025}. We remark that generalizations of Gaussian boson sampling to coherent inputs have been discussed as well~\cite{Thekkadath2022}. In this case, Eqs.~\eqref{eq:Complex Squeezer Parameters} and~\eqref{eq:Boson sampler simplification} continue to hold, but the initial coherent amplitude must be modified, see App.~\ref{subsec:CSq OTA Appendix}.

While no particular readout is preferred by the OTA, sampling the particle-number distribution experimentally provides access to the full information contained in the (real-space) particle-number sector of the simulated quantum field theory, which complements current approaches based on full-counting statistics~\cite{Bertini2023,Joshi2025}. From this, higher-order correlators and information measures can be inferred to probe symmetry-resolution during non-equilibrium dynamics~\cite{Ares2023,Rylands2024}. Therefore, even in the simplest scenario---vacuum inputs and linear dynamics---the OTA offers a pathway to quantum advantage across a wide range of physically relevant settings.

\section{Simulating quantum fields}
\label{sec:Simulation}
Next, we discuss some representative classes of quantum field theories that can be simulated using the OTA. For simplicity, we consider one spatial dimension $d=1$ and periodic boundary conditions. Generalizations to higher dimensions are straightforward: the total mode number scales as $\tilde{N}^d$, affecting only the interferometer's structure. Similarly, boundary conditions are arbitrary, see App.~\ref{subsec:FiniteSizeEffects} for a discussion of boundary effects.

\subsection{General Hamiltonian structure}
\label{subsec:HamiltonianStructure}
A central observation is that many scalar-field Hamiltonians become diagonal in momentum space. In the discretized picture, this is intimately related to the $\phi$-block of the Hamiltonian matrix $H$ corresponding to a circulant (and by definition, real and symmetric) matrix
\begin{equation}
    \begin{split}
        H^{\phi} &= \text{circ}(h^{\phi}_1, h^{\phi}_2, \dots, h^{\phi}_N) \\
        &= \begin{pmatrix}
            h^{\phi}_1 & h^{\phi}_N & \dots & h^{\phi}_3 & h^{\phi}_2 \\
            h^{\phi}_2 & h^{\phi}_1 & h^{\phi}_N & \dots & h^{\phi}_3 \\
            \vdots & h^{\phi}_2 & h^{\phi}_1 & \ddots & \vdots \\
            h^{\phi}_{N-1} & \dots & \ddots & \ddots & h^{\phi}_N \\
            h^{\phi}_N & h^{\phi}_{N-1} & \dots & h^{\phi}_2 & h^{\phi}_1
        \end{pmatrix},
    \end{split}
\end{equation}
with the symmetric constraint $h_j = h_{N+2-j}$ understood (such that it has only $\floor{N/2} + 1$ independent elements). 

Indeed, circulant matrices are diagonalized by a discrete Fourier transform (DFT). Hence, the required parameters for the phase shifters and squeezers, cf. Eqs.~\eqref{eq:OTAParameters1} and~\eqref{eq:OTAParameters2}, respectively, can be computed analytically in full generality. The real eigenvalues $\lambda_j^{\phi}$ of $H^{\phi}$ read
\begin{equation}
    \lambda_j^{\phi} = \sum_{k=1}^{N} h^{\phi}_k \cos \dfrac{2 \pi (j-1) (k-1)}{N},
    \label{eq:CirculantEigenvalues}
\end{equation}
and typically encode the dispersion of the theory $\lambda_j^{\phi} \propto \omega_j$ since often $\lambda_j^{\pi} = $ const.

To construct the corresponding set of \textit{real} eigenvectors $P_N$, we consider trigonometric functions instead of the usual complex-valued exponentials. The first eigenvalue $\lambda_1^{\phi} = \sum_{k=1}^N h_k$ has the constant eigenvector $p_1 = \frac{1}{\sqrt{N}} \left[1, \dots, 1 \right]^{\intercal}$. The eigenvalues in $2 \le j \le \floor{N/2}$ are two-fold degenerate with respect to reflection at $N/2 + 1$, i.e., $\lambda_j^{\phi} = \lambda_{N+2-j}^{\phi}$. Each pair of such eigenvalues comes with two real eigenvectors
\begin{equation}
    \begin{split}
        (p_j^{\text{cos}})_k &= \sqrt{\tfrac{2}{N}} \cos \tfrac{2\pi (j-1) (k-1)}{N}, \\
        (p_j^{\text{sin}})_k &= \sqrt{\tfrac{2}{N}} \sin \tfrac{2\pi (j-1) (k-1)}{N}.
    \end{split}
    \label{eq:DFTEigenvectors}
\end{equation}
For even $N$, there is an additional eigenvalue $\lambda_{N/2+1} = \sum_{k=1}^N h^{\phi}_k (-1)^{k-1}$ corresponding to the alternating eigenvector $p_{N/2+1} = \frac{1}{\sqrt{N}} \left[1, -1, \dots \right]^{\intercal}$. We emphasize that the eigenvectors are independent of the matrix elements $h_j^{\phi}$; hence, the interferometer layers are identical for all Hamiltonians diagonalized by a DFT, including time-dependent scenarios.

\subsection{Relativistic theory} 
\label{subsec:Relativistic Real Field}
The archetypal scalar field theory, that is, the relativistic, real, and massive scalar field, is characterized by the Lagrangian density
\begin{equation} \label{eq:Relativistic Real Lagrangian Density}
    \mathcal{L}_{\text{R}} = \dfrac{1}{2} \left[(\partial_{t} \phi)^{2} - (\partial_{x} \phi)^{2} - m^{2} \phi^{2} \right],
\end{equation}
where $m \ge 0$ denotes the mass. A special case is the massless limit $m \to 0$, which renders the theory conformal, and serves as an effective theory for various phenomena in the low-energy regime: phononic excitations of a Bose-Einstein condensate (BEC)~\cite{Pitaevskii2016}; the Tomonaga-Luttinger-Liquid after bosonization~\cite{Colemann1975} describing linear excitations of interacting fermions in, e.g., quantum wires~\cite{Blumenstein2011}; excitations of the XXZ Heisenberg spin chain at criticality~\cite{Haldane1981}. The massive variant appears, e.g., for the linear fluctuations between two strongly tunnel-coupled BECs~\cite{Tajik2023,Tajik2023b}.

After identifying the conjugate momentum field $\pi = \partial_t \phi$ and discretizing the theory as discussed in Sec.~\ref{sec:Preliminaries}, the Hamiltonian to be simulated reads
\begin{equation}
    \boldsymbol{H}_{\text{R}} = \dfrac{1}{2 \epsilon} \sum_{j \in \mathcal{J}} \left[ \boldsymbol{\pi}^{2}_{j} + (\boldsymbol{\phi}_{j+1} - \boldsymbol{\phi}_{j})^{2} + \epsilon^{2} m^{2} \boldsymbol{\phi}^{2}_{j} \right],
    \label{eq:Relativistic Real Hamiltonian Operator in Real Space}
\end{equation}
with $\epsilon > 0$ denoting the lattice spacing and $j \in \mathcal{J} = \{1, \ldots, N\}$ being a discrete label, such that $L = N \epsilon$. We refer to App.~\ref{subsubsec:OpenBoundaries} for a treatise of open boundary conditions.

It is evident that this Hamiltonian is of the form required for applying the OTA since the corresponding Hamiltonian matrix decomposes as $H_{\text{R}} = H_{\text{R}}^{\phi} \oplus H_{\text{R}}^{\pi}$ and the diagonal matrix $H_{\text{R}}^{\pi} = \mathds{1}_{N}/\epsilon$ commutes with any other matrix. Further, the $\phi$-block is indeed circulant as
\begin{equation} \label{eq:Relativistic Real Hamiltonian Matrix Phi-Block}
    H_{\text{R}}^{\phi}
    = \begin{pmatrix}
        \mu & \nu &  \ldots & 0 & \nu \\
        \nu & \mu &  \ldots & 0 & 0 \\
        \vdots & \ddots & \ddots & \ddots & \vdots \\
        0 & 0 &  \ldots &\mu & \nu \\
        \nu & 0 & \ldots & \nu & \mu
    \end{pmatrix} ,
\end{equation}
where we introduced $\mu = \epsilon m^{2} - 2 \nu$ and $\nu = - 1/\epsilon$.

Following Sec.~\ref{subsec:HamiltonianStructure}, $H_{\text{R}}^{\phi}$ is diagonalized by a DFT, and hence the phase shifter and squeezer parameters follow directly from~\eqref{eq:CirculantEigenvalues} and are given by $d_j = \omega_j, \gamma_j = \epsilon \, \omega_j$ with the discretized relativistic dispersion
\begin{equation} \label{eq:Dispersion relation real relativistic theory}
    \omega^2_{j} = m^{2} + \dfrac{4}{\epsilon^{2}} \sin^{2} \left[ \tfrac{ \pi (j-1)}{N} \right].
\end{equation}
In App.~\ref{sec:Complete decomposition N=5}, we exemplarily list all parameters for $N=5$. We remark that the massless limit $m \to 0$ causes the first pair of parameters to vanish $d_1 = \gamma_1 = 0$, independent of $\epsilon$ and $N$. At first glance, this would require infinite squeezing $z_1 \to \infty$ in the first mode. However, since the corresponding phase shift disappears as well, $\varphi_1 (t) = 0$, the two squeezer transformations cancel each other. Hence, any non-zero value can be chosen for $z_1$ without affecting the physics.

\subsection{Non-local theories}
\label{subsec:Fractional Laplacian Real Field}
With the relativistic scalar serving as our baseline, we investigate how far the OTA can stretch by altering relevant physical properties. A key aspect of a field theory is its coupling range. Many physical processes are inherently local in the sense that changing a physical quantity at a specific spatial position only affects its values in the vicinity of this point. In the Lagrangian, locality manifests as two spatial derivatives acting on the field in the continuum $\sim (\partial_x \phi)^2$ and nearest-neighbor coupling terms in the discrete setting $\sim (\phi_{j+1} - \phi_j)^2$, as is the case for the relativistic theory, cf. Eqs.~\eqref{eq:Relativistic Real Lagrangian Density} and~\eqref{eq:Relativistic Real Hamiltonian Operator in Real Space}, respectively.

In the following, we consider non-local theories, in particular, the class of long-range theories (see~\cite{Defenu2023} for a review), characterized by a power-law decay of couplings at large distances. The long-range dynamics of both non-interacting and interacting systems have been extensively studied in trapped-ion~\cite{Schachenmayer2013,Richerme2014,Jurcevic2014,Gaerttner2017,Zhang2017,Brydges2019,Joshi2020,Monroe2021} and Rydberg~\cite{Browaeys2020} setups as well as optical cavities~\cite{Periwal2021}, with varying control over the coupling range: Dipolar interactions are characteristic for Rydberg atoms and magnetic atoms or molecules, while trapped-ion systems can also realize all-to-all couplings.

To study long-range effects beyond the many-body regime, i.e., in a proper field-theoretic setting using the OTA, we consider the family of theories where the Laplace operator is raised to some positive power $\alpha > 0$, thereby controlling the range of the couplings. The Lagrangian reads
\begin{equation}
    \mathcal{L}_{\text{FL}} = \dfrac{1}{2} \left[ (\partial_{t} \phi)^{2} - \phi \, (-\partial^2_{x})^{\alpha/2} \phi - m^{\alpha} \phi^{2} \right],
    \label{eq:FractionalLaplacianLagrangian}
\end{equation}
and is referred to as fractional Laplacian theory~\cite{Nezhadhaghighi2013,Nezhadhaghighi2014,Rajabpour2015,Basa2020,Lischke2020,Daoud2022,Roy2022}. The discrete Hamiltonian evaluates to
\begin{equation}
    \boldsymbol{H}_{\text{FL}} = \dfrac{1}{2 \epsilon} \sum_{j \in \mathcal{J}} \Big[ \boldsymbol{\pi}^{2}_{j} + \sum_{j' \in \mathcal{J}} \boldsymbol{\phi}_j f_{j j'} (\alpha) \boldsymbol{\phi}_{j'} + \epsilon^{2} m^{\alpha} \boldsymbol{\phi}^{2}_{j} \Big],
    \label{eq:FractionalLaplacianHamiltonian}
\end{equation}
with $f_{j j'}$ being a circulant and symmetric matrix, which, in the infinite volume limit, is given by
\begin{equation}
    f_{j j'} (\alpha) = - \dfrac{\Gamma (-\tfrac{\alpha}{2} + j - j') \Gamma (\alpha + 1)}{\pi \Gamma (1 + \tfrac{\alpha}{2} + j - j')} \sin \left( \dfrac{\alpha \pi}{2} \right).
\end{equation}
We refer to App.~\ref{sec:Fractional Laplacian Theory in Momentum Space} for the corresponding expression on a finite-size lattice.

This model captures all essential features of non-local theories: The local couplings associated with the relativistic theory \eqref{eq:Relativistic Real Lagrangian Density} are obtained for $\alpha = 2$, in which case $f_{j j'} (2) = 2 \delta_{j j'} - \delta_{j j'+1} - \delta_{j j' + N}$. When $\alpha < 2$, the theory exhibits long-range couplings up to all-to-all couplings in the limit $\alpha \to 0$. Additionally, $\alpha > 2$ and $\alpha \in \mathbb{N}$ describe ultra-short-range couplings constituting higher-derivative theories that are often considered in the context of quantum gravity~\cite{Gibbons2019,Bambi2024}. 

The Hamiltonian matrix decomposes $H_{\text{FL}} = H_{\text{FL}}^{\phi} \oplus H_{\text{FL}}^{\pi}$ with circulant and symmetric $(H_{\text{FL}}^{\phi})_{j j'} = \epsilon m^{\alpha} \delta_{j j'} + f_{j j'} (\alpha)$ and $H_{\text{FL}}^{\pi} = \mathds{1}_N/\epsilon$. Hence, the OTA's parameters are again given by $d_j = \omega_j (\alpha), \gamma_j = \epsilon \, \omega_j (\alpha)$, with the fractional dispersion
\begin{equation}
    \omega^2_{j} (\alpha) = m^{\alpha} + \dfrac{2^{\alpha}}{\epsilon^{2}} \abs*{\sin^{\alpha} \left[ \tfrac{ \pi (j-1)}{N} \right]}.
    \label{eq:FractionalDispersion}
\end{equation}
We note that other types of non-local theories can also be simulated by the OTA, such as the exponentiated fractional Laplacian considered in Refs.~\cite {Shiba2014,Basa2020}, for which the $\pi$-component of the Hamiltonian matrix is also proportional to the identity.

\subsection{Complex fields and the free Bose gas}
\label{subsec:BoseGas}
When generalizing to complex-valued fields, the relativistic Lagrangian density attains the form
\begin{equation} \label{eq:Relativistic Complex Lagrangian Density}
    \mathcal{L}_{\text{C}} = (\partial_{t} \psi^{*}) (\partial_{t} \psi) - (\partial_{x} \psi^{*}) (\partial_{x} \psi) - m^{2} \psi^{*} \psi.
\end{equation}
This model describes particles and anti-particles with their particle number difference being a conserved charge corresponding to a $U(1)$ symmetry. After decomposing the complex field into two real-valued fields $\phi_{1}$ and $\phi_{2}$ via $\psi = (\phi_{1} + i \phi_{2})/\sqrt{2}$, the Lagrangian decouples $\mathcal{L}_{\text{C}} = \mathcal{L}_{\text{R},1} + \mathcal{L}_{\text{R},2}$. Hence, the complex relativistic theory can be quantum-simulated by doubling the number of modes.

Of particular interest is the non-relativistic limit in Eq.~\eqref{eq:Relativistic Complex Lagrangian Density}. To this end, we factor out fast-oscillating contributions by rescaling the field $\psi \to e^{-i(m - V)t} \psi/\sqrt{2m}$, where $V >0$ represents an external field. Then, neglecting second-order time derivatives, which amounts to the limit $c \to \infty$ after reinstating units, gives
\begin{equation}
    \mathcal{L}_{\text{NRC}} = i \psi^{*} \partial_{t} \psi - \dfrac{1}{2m} (\partial_{x} \psi^{*}) (\partial_{x} \psi) - V \psi^{*} \psi,
\end{equation}
which describes, for instance, a free Bose gas in an external trapping potential $V=V(x)$~\cite{Pitaevskii2016}.

For discretizing the non-relativistic theory, we parametrize the complex field in terms of the two canonical fields as $\boldsymbol{\psi} = (\boldsymbol{\phi} + i \boldsymbol{\pi})/\sqrt{2\epsilon}$, with the additional factor $1/\sqrt{\epsilon}$ guaranteeing that \eqref{eq:CCRStandard} is fulfilled. This results in the decoupled Hamiltonian
\begin{equation}
    \begin{split}
        \hspace{-0.1cm}\boldsymbol{H}_{\text{NR}} \hspace{-0.05cm}= \hspace{-0.05cm}\frac{1}{2} \hspace{-0.05cm}\sum_{j \in \mathcal{J}} \hspace{-0.05cm}\Big\{ &\frac{1}{2m\epsilon^2} \left[ (\boldsymbol{\phi}_{j+1} \hspace{-0.05cm}-\hspace{-0.05cm}\boldsymbol{\phi}_{j})^2 \hspace{-0.05cm}+\hspace{-0.05cm} (\boldsymbol{\pi}_{j+1}\hspace{-0.05cm}-\hspace{-0.05cm}\boldsymbol{\pi}_{j})^2 \right] \\
        &+ V_j \left( \boldsymbol{\phi}_j^2 + \boldsymbol{\pi}_j^2 \right) \Big\}.
    \end{split}
    \label{eq:NonrelativisticHamiltonian}
\end{equation}
After inverting $\boldsymbol{a}_j = (\boldsymbol{\phi}_j+i\boldsymbol{\pi}_j)/\sqrt{2}$, the latter attains the form of a non-interacting Bose-Hubbard model~\cite{Jaksch1998,Greiner2002}
\begin{equation}
    \begin{split}
        \boldsymbol{H}_{\text{BH}} = \frac{1}{2} \sum_{j \in \mathcal{J}} \Big[&- \dfrac{1}{m\epsilon^2} \left(\boldsymbol{a}^{\dagger}_{j+1} \boldsymbol{a}_{j} + \boldsymbol{a}^{\dagger}_{j} \boldsymbol{a}_{j+1} \right) \\
        &+ 2 \Big(V_j + \dfrac{1}{m\epsilon^2} \Big) \boldsymbol{a}^{\dagger}_{j} \boldsymbol{a}_{j} \Big].
    \end{split}
    \label{eq:BoseHubbardModel}
\end{equation}
describing ultracold atoms hopping between neighboring sites in an optical lattice, with hopping amplitude $-1/(m\epsilon^2)$ and a site-dependent potential $2[V_j + 1/(m\epsilon^2)]$. The corresponding  Hamiltonian matrix decomposes as $H_{\text{NR}} = H^{\phi}_{\text{NR}} \oplus H^{\pi}_{\text{NR}}$, with $H^{\phi}_{\text{NR}} = H^{\pi}_{\text{NR}}$. Since the two blocks agree, we find $\gamma_j = 1$, such that no squeezing is required. Thus, the non-relativistic theory, or equivalently, the Bose-Hubbard model, can be simulated with passive linear optics only. 

A general trapping profile $V_j$ breaks the circulanticity of both blocks of $H$, which read
\begin{equation}
    H_{\text{NR}}^{\phi}
    = H_{\text{NR}}^{\pi} = \begin{pmatrix}
            \mu_{1} & \nu & \dots & 0 & \nu \\
            \nu & \mu_{2} & \nu & \dots & 0 \\
            \vdots & \nu & \mu_{3} & \ddots & \vdots \\
            0 & \dots & \ddots & \ddots & \nu \\
            \nu & 0 & \dots & \nu & \mu_{N}
        \end{pmatrix},
        \label{eq:NonCirculantHamiltonian}
\end{equation}
with $\mu_j = V_j - 2 \nu$ and $\nu = -1/(2 m \epsilon^2)$. There exists no closed-form analytic solution for the eigenvalues and eigenvectors of the latter matrix in general; hence, numerical methods must be used to determine the circuit parameters. When the potential is constant $V \equiv V_j$, the parameter $\mu \equiv \mu_j = V - 2 \nu$ amounts to the chemical potential, and the blocks of the Hamiltonian matrix reduce to the relativistic form~\eqref{eq:Relativistic Real Hamiltonian Matrix Phi-Block}. In this case, we obtain $d_j = \omega_j$ for the phase shifts, with the non-relativistic dispersion
\begin{equation}
    \omega_j = V + \frac{2}{m \epsilon^2} \sin^{2} \left[ \tfrac{ \pi (j-1)}{N} \right].
\end{equation}

\subsection{Curved spacetimes}
Since more than two decades, simulating quantum fields in curved spacetime~\cite{Birrell1982,Mukhanov2007} is of substantial interest to experimentally probe phenomena that remain inaccessible in the nightsky, including cosmological particle production~\cite{Jain2007,Eckel2018,Steinhauer2021,Haas2022b,Haas2022c,Haas2022d}, the Hawking~\cite{Garay2000,Philbin2008,Steinhauer2014,MunozDeNova2019} and Unruh~\cite{Hu2019} effects, and rotating analog black holes~\cite{Svancara2024}. The standard approach in the field of analog gravity is to consider low-energy perturbations in a fluid, whose effective equation of motion is that of a massless Klein-Gordon field in a curved spacetime dictated by the background fluid flow in the form of the so-called acoustic metric~\cite{Barcelo2011,Braunstein2023}. It is well known, however, that this one-to-one correspondence comes with several restrictions. First, the acoustic metric has a particular form. Hence, it is not possible to engineer arbitrary spacetimes this way. Second, the mapping breaks down in $1+1$ dimensions, as conformal symmetry prevents the wave equation for the perturbations from being rewritten in terms of a proper metric---thereby limiting the majority of experiments to qualitative analogies~\cite{Eckel2018,Steinhauer2021}. Third, simulating \textit{massive} fields typically requires two-component fluids~\cite{Weinfurtner2007}, which has not yet been demonstrated. Fourth, the standard observable in fluid-based analog gravity experiments, that is, the two-point correlation function of the density contrast~\cite{Steinhauer2014,MunozDeNova2019,Haas2022b,Haas2022c}, is, to first order, related to the two-point correlator of the conjugate momentum field $\boldsymbol{\pi} (x)$ only. Hence, state reconstruction in a Gaussian setting via estimating the covariance matrix is not possible in a direct way.

In what follows, we show that a photonic simulation with the OTA resolves all four problems and thus, provides a substantially more versatile approach to simulating quantum fields in curved spacetime compared to previous methods. As for fluid-based analogies, we do not include back-reaction effects on the underlying metric.

We start again from the relativistic theory~\eqref{eq:Relativistic Real Lagrangian Density} and generalize to arbitrary spacetime geometries by introducing a metric $g_{\mu \nu}$ in the Lagrangian density,
\begin{equation}
    \mathcal{L}_{\text{CS}} = \dfrac{1}{2} \sqrt{-g} \left[ g^{\mu \nu} (\partial_{\mu} \phi) (\partial_{\nu} \phi ) - m^{2} \phi^{2} \right].
    \label{eq:CurvedSpacetimeLagrangian}
\end{equation}
Here, $x^{\mu} = (t,x)$ labels spacetime coordinates and $\sqrt{-g} = \sqrt{- \text{det}(g_{\mu \nu})}$ renders the volume element invariant under arbitrary coordinate transformations. 

A remarkable result from differential geometry is that \textit{all} two-dimensional manifolds are conformally flat~\cite{Misner2017}. In general, a pseudo-Riemannian manifold is conformally flat if for each point there exists an open neighborhood that can be mapped to flat, i.e., Minkowski space, via a conformal transformation. These transformations preserve angles but scale lengths locally; the stereographic projection from the $2$-sphere to the plane is a prominent example. Intuitively, conformal flatness amounts to spacetime regions appearing scaled, but never sheared, with respect to flat space. While conformally flat manifolds exist in arbitrary dimensions, two-dimensional manifolds are always conformally flat, which is a consequence of low dimensionality: The metric is symmetric by definition and hence has three degrees of freedom in two dimensions. Two of them single out spacetime coordinates, while the remaining one expresses local rescalings. Hence, the metric is necessarily of the form
\begin{equation} \label{eq:Conformal factor}
    g_{\mu \nu}  = f (x^\mu) \, \eta_{\mu \nu} = f (x^{\mu}) \, \text{diag}(1,-1),
\end{equation}
where $f(x^{\mu})>0$ denotes the conformal factor and $\eta_{\mu \nu} = \text{diag} (1,-1)$ is the Minkowski metric. The high degree of symmetry is also reflected in the Riemann tensor, which quantifies curvature. In two dimensions, it has only \textit{one} independent component, i.e., $R_{\mu \nu \alpha \beta} = (R/2) (g_{\mu \alpha} g_{\nu \beta} - g_{\mu \beta} g_{\nu \alpha})$, showing that spacetime curvature is entirely determined by the Ricci scalar. Expressed in terms of the conformal factor, the latter reads $R=(1/f) \eta^{\mu \nu} \partial_{\mu} \partial_{\nu} \ln f$. 

By exploiting conformal flatness, the corresponding discretized Hamiltonian attains the simple form
\begin{equation}
    \boldsymbol{H}_{\text{CS}} = \dfrac{1}{2 \epsilon} \sum_{j \in \mathcal{J}} \bigg[ \boldsymbol{\pi}_{j}^{2} + (\boldsymbol{\phi}_{j+1} - \boldsymbol{\phi}_{j})^{2} + \epsilon^{2} m_{\text{eff}}^{2}(t,\epsilon j)  \boldsymbol{\phi}_{j}^{2} \bigg],
    \label{eq:CurvedSpacetimeHamiltonian}
\end{equation}
which is reminiscent of~\eqref{eq:Relativistic Real Hamiltonian Operator in Real Space} with a spacetime-dependent effective mass $m_{\text{eff}}(x^{\mu}) = m \sqrt{f(x^{\mu})}$. In general, this Hamiltonian breaks the circulanticity of the $\phi$-block, which is of the form~\eqref{eq:NonCirculantHamiltonian} with $\mu_{j} (t) = \epsilon m^{2} f(t,\epsilon j) - 2 \nu$ and $ \nu = - 1/\epsilon$, while $H^{\pi} = \mathds{1}_N/\epsilon$. 

In what follows, we explore spacetime metrics prominent in general relativity. In particular, we provide conformal coordinates favorable for simulation via the OTA, along with the corresponding conformal factors and expressions for the Ricci scalar. We stress, however, that, more generally, coordinates can be chosen to probe the aspects of interest.

\subsubsection{Variations in time: Cosmology}
Let us first consider time-dependent scenarios $f = f (t)$ which are relevant for cosmology. When imposing periodic boundary conditions, the $\phi$-block of~\eqref{eq:CurvedSpacetimeHamiltonian} boils down to the relativistic result \eqref{eq:Relativistic Real Hamiltonian Matrix Phi-Block} with time-dependent diagonal entries $\mu (t) = \epsilon m^2 f(t) - 2 \nu$, thus recovering circulanticity, leading to a now time-dependent relativistic dispersion
\begin{equation}
    \omega^2_{j}(t) = m^{2} f(t) + \dfrac{4}{\epsilon^{2}} \sin^{2} \left[ \tfrac{ \pi (j-1)}{N} \right].
    \label{eq:Dispersion relation real relativistic theory on curved spacetime}
\end{equation}
Since the diagonalizing DFT remains time-independent, we can apply the time-dependent OTA (Sec.~\ref{subsubsec:Time-dependent OTA}) to simulate cosmological dynamics.

On large scales, our universe is homogenous and isotropic, which is captured by the Friedmann–Lemaître–Robertson–Walker (FLRW) metric $g_{\mu \nu} = \text{diag}[1, -a^2(t)]$~\cite{Weinberg2008}. Therein, $a(t)$ specifies the scale factor, which models cosmological expansion and contraction as a function of cosmic time $t$. It is convenient to introduce conformal time $\eta (t) = \int_{0}^t \mathrm{d}t'/a(t')$~\cite{Weinberg2008}. Indeed, when identifying laboratory with conformal time, i.e., considering coordinates $x^{\mu} = (\eta, x)$, the FLRW metric attains its conformally flat form~\eqref{eq:Conformal factor} with conformal factor
\begin{equation}
    f_{\text{FLRW}}(\eta) = a^2 (\eta).
\end{equation}
Spacetime curvature manifests in the Ricci scalar as $R=(2/a^3) a'' - (2/a^4) a'^2$, where $a' (\eta) \equiv \mathrm{d} a(\eta)/\mathrm{d} \eta$.

Hence, when operating the time-dependent OTA in conformal time, cosmological situations with \textit{arbitrary} scale factors $a(\eta)$ can be engineered. This goes substantially beyond ultracold atom simulations, where only power-law scale-factors have been considered \cite{Haas2022b,Haas2022c}. Cosmologically relevant choices for $a(\eta)$ are summarized in Tab.~\ref{tab:Cosmology}, see also Refs.~\cite{Birrell1982,Mukhanov2007}.

\subsubsection{Variations in space: Horizons and (Anti-) de Sitter spaces}
Spatial variations break the circulanticity of the $\phi$-block. Yet, time-independent scenarios are always OTA-simulatable, with circuit parameters that must be computed numerically in general. In contrast to cosmology, the coordinates in which the metric assumes its conformally flat form depend strongly on the spacetime of interest; see Tab.~\ref{tab:SpatialVariations} for an overview and Refs.~\cite{Birrell1982,Mukhanov2007}.

Of particular interest are (Killing) horizons that causally disconnect a spacetime region from its surroundings. The simplest example is the so-called Rindler spacetime, which describes flat spacetime as perceived by an observer moving at constant acceleration, giving rise to thermal Unruh radiation. Starting from flat-space coordinates $(t,x)$, the accelerated observer uses Rindler tortoise coordinates $(\eta, \rho)$ defined via $t = (e^{\kappa \rho}/\kappa) \sinh (\kappa \eta)$ and $x = (e^{\kappa \rho}/\kappa) \cosh (\kappa \eta)$ covering the right Rindler wedge $x > \abs{t}$. 

The prime examples of horizons are black-hole-like metrics of the form $g_{\mu \nu} = \text{diag} [f(r), -1/f(r)]$, where $r$ denotes a radial coordinate and spacetime curvature is given by the simple formula $R=-\mathrm{d}^2 f(r)/\mathrm{d}r^2$. The horizon sits at $f(r_{\text{H}}) = 0$. As in the Rindler setup, observers near the horizon observe a thermal spectrum---Hawking radiation. Analogously to conformal time, we introduce tortoise coordinates $r^* (r) = \int_{0}^r \mathrm{d}r'/f(r')$ to obtain the conformally flat form, in which case $f(r^*)$ becomes the conformal factor. Note here that one needs to solve the differential equation $r^* = r^*(r)$ to find $f(r^*)$. This class encompasses the famous Schwarzschild spacetime $f(r)=1-r_s/r$, where $r_s = 2 M$ denotes the Schwarzschild radius in natural units for a black hole of mass $M$. Another prominent variant is the Reissner-Nordström metric for which $f(r) = 1-r_s/r+(Q/r)^2$, describing a black hole of mass $M$ and charge $Q$. We stress that in practice, black-hole-like metrics must feature a cutoff at finite $r>0$ to avoid the singularity at $r=0$.

\renewcommand{\arraystretch}{1.3}
\begin{table}[t!]
    \centering
    \setlength{\tabcolsep}{0pt}
    \begin{tabular}{p{2cm} p{2cm} p{2cm} p{2cm}}
    \toprule
    Type & $a(t)$ & $a(\eta)$ & $R (\eta)$ \\
    \midrule
    \rowcolor{Gray}
    \, Power law & $\propto t^q, q \neq 1$ & $\propto \abs{\eta}^{\frac{q}{1-q}}$ & $\propto \tfrac{2q}{q-1} (a \eta)^{-2}$ \\
    \multicolumn{4}{p{8cm}}{\footnotesize{Tunable expansion acceleration: $0 \le q < 1$ and $q>1$ describe de- and accelerated expansion, respectively; Includes radiation ($q=1/2$) and matter ($q=2/3$) dominated phases of our universe.}} \\
    \rowcolor{Gray}
    \, Milne & $\propto t$ & $\propto e^{\eta}$ & $0$ \\
    \multicolumn{4}{p{8cm}}{\footnotesize{Coasting expansion with vanishing curvature.}} \\
    \rowcolor{Gray}
    \, De Sitter & $\propto e^{H t}$ & $\propto - (H \eta)^{-1}$ & $\propto 2 H^2$ \\
    \multicolumn{4}{p{8cm}}{\footnotesize{Describes the rapid inflatory phase of our universe with Hubble rate $H$ and constant positive, i.e., spherical, curvature; Corresponds to the time-like patch of dS$_2$.}} \\
    \rowcolor{Gray}
    \, Bouncing & $\propto \sqrt{1 + t^2}$ & $\propto \cosh \eta$ & $\propto \text{sech}^4 \eta$ \\
    \multicolumn{4}{p{8cm}}{\footnotesize{Toy model avoiding the Big Bang singularity as $R(0)=1$ remains finite.}} \\
    \bottomrule
    \end{tabular}
    \caption{Scale factors relevant for cosmology.}
    \label{tab:Cosmology}
\end{table}
\renewcommand{\arraystretch}{1}

Interestingly, space-like (static) patches of (Anti-)de Sitter spacetimes, which correspond to constant curvature, i.e., maximally symmetric, spaces, also fall into the black-hole class: $f(r) \propto 1 \pm H^2 r^2$, with $- (+)$ for (Anti-)de Sitter metrics and $H>0$. It is well known that Anti-de Sitter spaces are central to holography~\cite{Ammon2015}, underscoring the prospects of the OTA for studying quantum gravity models once gravitational degrees of freedom are included. We remark that there are also alternative coordinate choices. This includes the conformal factor $f(r) \propto 1/(Hr)^2$ for AdS$_2$, describing the so-called Poincaré patch, which is conformally flat by construction.

As in cosmological scenarios, the OTA enables analog gravity experiments with spatially dependent metrics far beyond what is achievable with fluid-based approaches~\cite{Barcelo2011,Braunstein2023}. In the latter case, ultracold-atom experiments are restricted to toy models in which horizons arise from interpolating between over- and underdense regions via, for instance, $f(x) \propto 1 + \tanh(x)$~\cite{MunozDeNova2019}. 

\section{Information dynamics}
\label{sec:Applications}
As an application, we study the spreading of entanglement in spacetime after a global quantum quench for varying coupling range. We focus on identifying a regime in which continuum features emerge while keeping the number of modes at an experimentally feasible level. Given the classical hardness of probing the particle-number sector, we focus on Gaussian observables.

\renewcommand{\arraystretch}{1.3}
\begin{table}[t!]
    \centering
    \setlength{\tabcolsep}{0pt}
    \begin{tabular}{p{2.5cm} p{2cm} p{2.2cm} p{1.3cm}}
    \toprule
    Type & Coordinates & $f$ & $R$ \\
    \midrule
    \rowcolor{Gray}
    \, Rindler & $(\eta, \rho)$ & $e^{2 \kappa \rho}$ & $0$ \\
    \multicolumn{4}{p{8cm}}{\footnotesize{Describes a constantly accelerating observer in flat spacetime experiencing Unruh radiation.}} \\
    \rowcolor{Gray}
    \, Black-hole-like & $(t, r^*)$ & $f=f(r^*)$ & $-\tfrac{\mathrm{d}^2 f(r)}{\mathrm{d}r^2}$ \\
    \multicolumn{4}{p{8cm}}{\footnotesize{Includes, e.g., Schwarzschild $f(r) = 1-r_s/r$ and Reissner-Nordström $f(r) = 1-r_s/r+(Q/r)^2$ solutions featuring event horizons and Hawking radiation.}} \\
    \rowcolor{Gray}
    \, De Sitter & $(t, r^*)$ & $\propto \text{sech}^2 (H r^*)$ & $\propto 2H^2$ \\
    \multicolumn{4}{p{8cm}}{\footnotesize{Static patch of a maximally symmetric space with constant positive curvature and $H>0$.}} \\
    \rowcolor{Gray}
    \, Anti-De Sitter & $(t, r^*)$ & $\propto \text{sec}^2 (H r^*)$ & $\propto -2H^2$ \\
    \multicolumn{4}{p{8cm}}{\footnotesize{Static patch of a maximally symmetric space with constant negative curvature; Relevant for holographic dualities.}} \\
    \bottomrule
    \end{tabular}
    \caption{Conformal factors encoding spatially-varying metrics.}
    \label{tab:SpatialVariations}
\end{table}
\renewcommand{\arraystretch}{1}

\subsection{Quench dynamics}
\label{subsec:Quench dynamics}
We consider the experimentally friendly vacuum state $\ket{0}$ as the input state, in which case the OTA reduces to processing squeezed vacuum inputs with linear optics, as shown in Fig.~\ref{fig:Boson sampler}. The input state describes zero photons, which is the ground state of the pre-quench Hamiltonian
\begin{equation}
    \boldsymbol{H}_{0}
    = \dfrac{1}{2 \epsilon} \sum_{j \in \mathcal{J}}
    ( \boldsymbol{\pi}^{2}_{j} + \boldsymbol{\phi}^{2}_{j} ).
    \label{eq:PreQuenchHamiltonian}
\end{equation}
Importantly, $\ket{0}$ belongs to the class of Gaussian states, which are characterized by their mean and covariance 
\begin{equation}\label{eq:Definition of Mean and Covariance}
    r = \Tr \{ \boldsymbol{\rho} \boldsymbol{r} \}, \quad \sigma_{j j'} = \dfrac{1}{2} \Tr \left\{ \boldsymbol{\rho} \{ \boldsymbol{r}_j - r_j, \boldsymbol{r}_{j'} - r_{j'} \} \right\},
\end{equation}
respectively, with the latter encoding all two-point correlation functions. The ground state of~\eqref{eq:PreQuenchHamiltonian} is Gaussian with $r_0=0$ and $\sigma_0 = (1/2)\mathds{1}_{2N}$.

We time-evolve the state under the relativistic theory~\eqref{eq:Relativistic Real Hamiltonian Operator in Real Space}, which constitutes a quantum quench of both the mass and the couplings at $t=0$. We remark that the often-considered mass quench would instead require the input state to be the ground state of the Hamiltonian~\eqref{eq:Relativistic Real Hamiltonian Operator in Real Space} with a different mass---thus describing no quasi-particles (with respect to the initial theory) instead. In the relativistic case, this would necessitate additional squeezers during state preparation, see e.g., Ref.~\cite{Marshall2015}.

During time evolution, the state remains Gaussian, with $r(t) = S(t) r_0 = 0$ and $\sigma (t) = S(t) \sigma S^{\intercal} (t)$~\cite{Weedbrook2012}. This greatly simplifies the study of quantum entanglement and correlations, as standard quantum information measures can be expressed in terms of the covariance matrix only. From an experimental point of view, the covariance matrix can be reconstructed through homodyne measurements~\cite{DAuria2009}, with the number of measurement settings being proportional to (sub-)system size. Multimode homodyne measurements, also relevant to continuous-variable quantum computing and cluster state generation~\cite{Ferrini2013}, have been demonstrated on various experimental platforms~\cite{Armstrong2012,Harms2014,Cai2017,Jia2025,AghaeeRad2025}.

\subsection{Benchmarks: relativistic fields}
\label{subsec:Dynamical entanglement structure}

\subsubsection{Entanglement generation}
\begin{figure}[t]
    \centering
    \includegraphics[width=0.99\columnwidth]{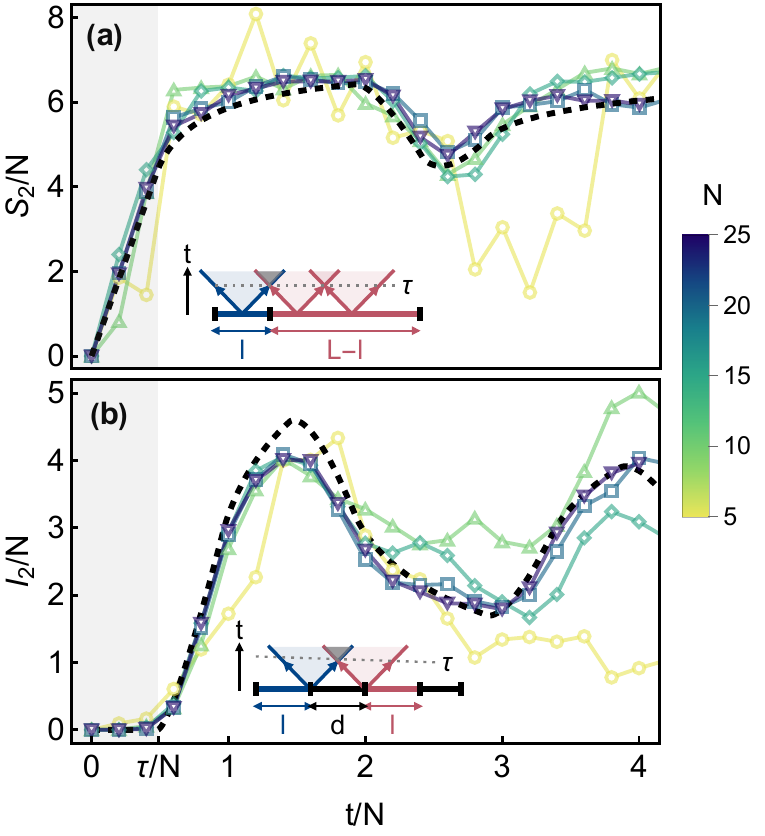}
    \caption{\textbf{(a)} Time evolution of the Rényi-2 entanglement entropy~\eqref{eq:Renyi2Entropy} (scaled by $10^2$) of the subregion $[0,\ell]$ following a quench from the optical vacuum~\eqref{eq:PreQuenchHamiltonian} to the relativistic theory~\eqref{eq:Relativistic Real Hamiltonian Operator in Real Space} with $\ell=L/5, m=1, \epsilon=2$ for varying $L=N \epsilon$ against the field-theoretic prediction~\eqref{eq:Renyi2EntropyQFT} (black dashed curve). The light-cone-like motion of quasi-particles generates entanglement (see inset), leading to a linear increase in entanglement entropy until the time $\tau/N = \ell/(2Nv_{\text{max}}) \approx 0.48$ (gray box) that the fastest quasi-particle pairs require to end up in distinct regions. \textbf{(b)} Same analysis for the Rényi-2 mutual information~\eqref{eq:Renyi2MI} between two spatial regions separated by a distance $d=\ell$. Correlations can only emerge after $t \ge \tau$, i.e., when the fastest quasi-particles have bridged the distance $d$. We refer to Fig.~\ref{fig:EntanglementGenerationOBC} in App.~\ref{subsubsec:OpenBoundaries} for a treatise of open boundaries.}
    \label{fig:EntanglementGeneration}
\end{figure}

We consider a spatial subregion $[0,\ell]$ with $\ell \le L$, see inset of Fig.~\hyperref[fig:EntanglementGeneration]{4(a)}. As the modes are coupled, the degrees of freedom in $[0,\ell]$ become entangled with the ones in the complementary region $(\ell,L]$. Thus, while the global state $\boldsymbol{\rho} (t)$ remains pure during the unitary evolution, the local state $\boldsymbol{\rho} (t,l) = \Tr_{L-\ell}\{ \boldsymbol{\rho} (t) \}$ follows the evolution of an open system described by a completely positive trace-preserving map~\cite{Pechukas1994}. We assess entanglement generation via the Rényi-2 entanglement entropy, which, for Gaussian states, attains the simple form
\begin{equation}
    S_{2} (t, \ell) = - \ln \Tr \{ \boldsymbol{\rho}^{2}(t, \ell) \} = \dfrac{1}{2} \ln \det [2 \sigma (t, \ell) ],
    \label{eq:Renyi2Entropy}
\end{equation}
where $\sigma (t,\ell)$ denotes the covariance associated with the local state $\boldsymbol{\rho} (t, \ell)$. Given homodyne samples, $S_{2} (t, \ell)$ can be estimated using the unbiased estimators discussed in~\cite{Ahmed1989} and refined in~\cite{Cai2015}. Their variance decays as $\mathcal{O}(N/\mathcal{N}_{\text{S}})$ with sample size $\mathcal{N}_{\text{S}}$ in the regime typical for photonic setups, that is, $\mathcal{N}_{\text{S}} \gg N$. Further, we remark that, for Gaussian states,~\eqref{eq:Renyi2Entropy} is equivalent to the recently introduced subtracted Wigner entropy~\cite{Haas2024c,Haas2024d,Haas2025a}.

Our main interest is to benchmark our quantum field simulator against QFT predictions. In the field theory context, the non-equilibrium dynamics of integrable models is understood in terms of a quasi-particle description~\cite{Calabrese2005,Calabrese2007,Calabrese2016,Calabrese2018,Alba2018,Calabrese2020,Modak2020} [see inset in Fig.~\hyperref[fig:EntanglementGeneration]{4(a)}]. Within this framework, the input state at $t=0$ acts as a source of quasi-particle pairs---momentum modes of opposite momenta $p$ and $-p$ ---which ballistically spread correlations throughout the system after the quench. Then, at any given time $t>0$, a pair contributes to the entanglement entropy if one particle ends up in the subregion of interest, while the other resides in the complementary region. Combining this with the fact that the non-equilibrium steady-state for $t \to \infty$ attains the form of a generalized Gibbs ensemble (GGE) \textit{locally}, i.e., $\boldsymbol{\rho}(\infty,l) \propto \text{exp}(-\sum_p y_p \boldsymbol{n}_p)$, where $\boldsymbol{n}_p = \tilde{\boldsymbol{a}}^{\dagger}_p \tilde{\boldsymbol{a}}_p$ counts (conserved) quasi-particles with momenta $p$ and $y_p$ are Lagrange multipliers~\cite{Langen2015,Vidmar2016}, leads to the QFT prediction in the infinite volume limit
\begin{equation}
    S_2 (t, \ell) = \int \dfrac{\mathrm{d} p}{2 \pi} s_2 (p) \min (2|v(p)|t, \ell).
    \label{eq:Renyi2EntropyQFT}
\end{equation}
Therein, $v(p) = \mathrm{d} \omega(p)/\mathrm{d}p = \sin(\epsilon p) /(\epsilon \omega(p))$ denotes the group velocity of the bosonic excitations and $s_2 (p)$ is the thermodynamic Rényi-2 entropy density associated with the GGE. We refer to App.~\ref{subsec:FiniteSizeEffects} for finite-size corrections for periodic and open boundaries, to App.~\ref{subsec:Quantum quench population} for an analytic calculation of $s_2 (p)$, and to App.~\ref{subsec:Divergences} for a comment on divergences appearing in the continuum limit $\epsilon \to 0$.

We compare the photonic simulation of the quench dynamics with the QFT prediction~\eqref{eq:Renyi2EntropyQFT} (black dashed curve) in Fig.~\hyperref[fig:EntanglementGeneration]{4(a)} for $\ell = L/5, m=1, \epsilon=2$ and varying system sizes $N = 5 - 25$ (see App.~\ref{sec:Complete decomposition N=5} for circuit parameters when $N=5$). Shortly after the quench, we observe a linear growth of the Rényi-$2$ entanglement entropy $S_2 (t,l) \propto t$ up to $t \le \tau = \ell /(2 v_{\text{max}})$, see gray region. The time $\tau$ marks the moment where two quasi-particles moving towards each other with the maximum velocity $v_{\text{max}}$ allowed by the relativistic dispersion have traveled a distance $\ell$ (see overlapping light cones in the inset).

After $t \ge \tau$, also the slower quasi-particles start to contribute to the entanglement between the subregion $[0,\ell]$ and its complement, but the growth of the entanglement entropy slows down due to saturation effects. For an infinitely extended system, the entropy would settle to its stationary value dictated by the GGE. For a finite system with periodic boundaries, however, the quasi-particles return to their initial positions every time they have traversed the system. This leads to so-called revivals in the entanglement entropy, marked by a temporary dip around the time $L/(2 v_{\text{max}}) = 5 \tau$ the fastest quasi-particles took to cross half of the system. As the quasi-particles' momenta vary, the entanglement is never entirely destroyed, and the dip attains a finite width. 

Over the whole time range considered, the photonic simulation is in remarkable agreement with the QFT prediction---even for small system sizes $N \sim 10$~\cite{Madsen2022,AghaeeRad2025}. Both the initial linear scaling and the revival dip(s) are particularly robust features that are clearly visible for even smaller mode numbers. This also holds true for open boundary conditions, in which case circulanticity is broken, see App.~\ref{subsubsec:OpenBoundaries}. Importantly, the circuit parameters fall into experimentally accessible regimes, see App.~\ref{sec:Complete decomposition N=5}. In general, implementing strong squeezing is most challenging. Here, the maximum squeezing required $z_{\text{max}}$, which corresponds to the zero-momentum mode, is independent of system size and evaluates to $\abs{z_{\text{max}}} \approx 0.5 \approx 4.3$dB---far below current records around $15$dB~\cite{Vahlbruch2016}. This demonstrates that photonic simulations of entanglement dynamics in quantum fields are well within the reach of current experimental technologies.

\subsubsection{Correlation spreading}
We now study how two equally sized spatial regions of length $\ell$ separated by a distance $d > 0$ become correlated after the quench, as shown in the inset of Fig.~\hyperref[fig:EntanglementGeneration]{4(b)}. To this end, we consider the Gaussian Rényi-$2$ mutual information
\begin{equation}
    \begin{split}
        I_{2}(t, \ell, d) &= 2 S_{2}(t, \ell) - S_{2}(t, \ell \cup \ell) \\
        &= \dfrac{1}{2} \ln \frac{\det \sigma^2 (t, \ell)}{\det \sigma (t, \ell \cup \ell)},
        \label{eq:Renyi2MI}
    \end{split}
\end{equation}
where $S_{2}(t, \ell \cup \ell)$ and $\sigma (t, \ell \cup \ell)$ are associated with the combined region $[0,\ell] \cup [\ell + d, 2 \ell + d]$. Analogously to the Rényi-$2$ entanglement entropy, the latter can be estimated using standard techniques~\cite{Ahmed1989,Cai2015} and equals the Wigner mutual information for Gaussian states~\cite{Haas2024c,Haas2024d,Haas2025a}.

In the quasi-particle picture and the infinite volume limit, the relativistic QFT~\eqref{eq:Relativistic Real Lagrangian Density} predicts
\begin{equation}
    \begin{split}
        I_{2}(t, \ell, d) & = \int \dfrac{\mathrm{d} p}{2 \pi} s_2 (p) \Big[ \max \left( 2|v(p)|t, d \right) \\
        &\hspace{2.2cm} + \max \left( 2|v(p)|t, d + 2\ell \right) \\
        &\hspace{2.2cm} - 2 \max \left( 2|v(p)|t, d + \ell \right) \Big],
    \end{split}
\end{equation}
see App.~\ref{subsec:FiniteSizeEffects} for finite-size corrections.

We benchmark the photonic simulation in Fig.~\hyperref[fig:EntanglementGeneration]{4(b)} for $d=L/5$, with all other parameters left unchanged. The two spatial regions become correlated only after the light cones of the fastest quasi-particles intersect, i.e., for $t \ge \tau$. We remark that this is the essence of Lieb-Robinson bounds~\cite{Lieb1972,Chenau2012,Tran2021,Chen2023}. Further, correlations decrease around the revival time $5 \tau$, with the dip's width being broadened compared to the Rényi-2 entanglement entropy as a result of having more possibilities for the quasi-particle configuration to resemble the initial state.

Strikingly, the early period $t \lesssim 3 \tau$ is well captured by simulations with small system sizes $N \sim 5 - 10$. We find deviations from the QFT prediction around the revival time. Here, reasonable agreement requires around $N \sim 20$ modes, which is still accessible on existing platforms. Both findings extend to open boundaries, see App.~\ref{subsubsec:OpenBoundaries}.

\subsection{Systematic study: Bending of light cones}

\begin{figure*}[t!]
    \centering
    \includegraphics[width=0.99\textwidth]{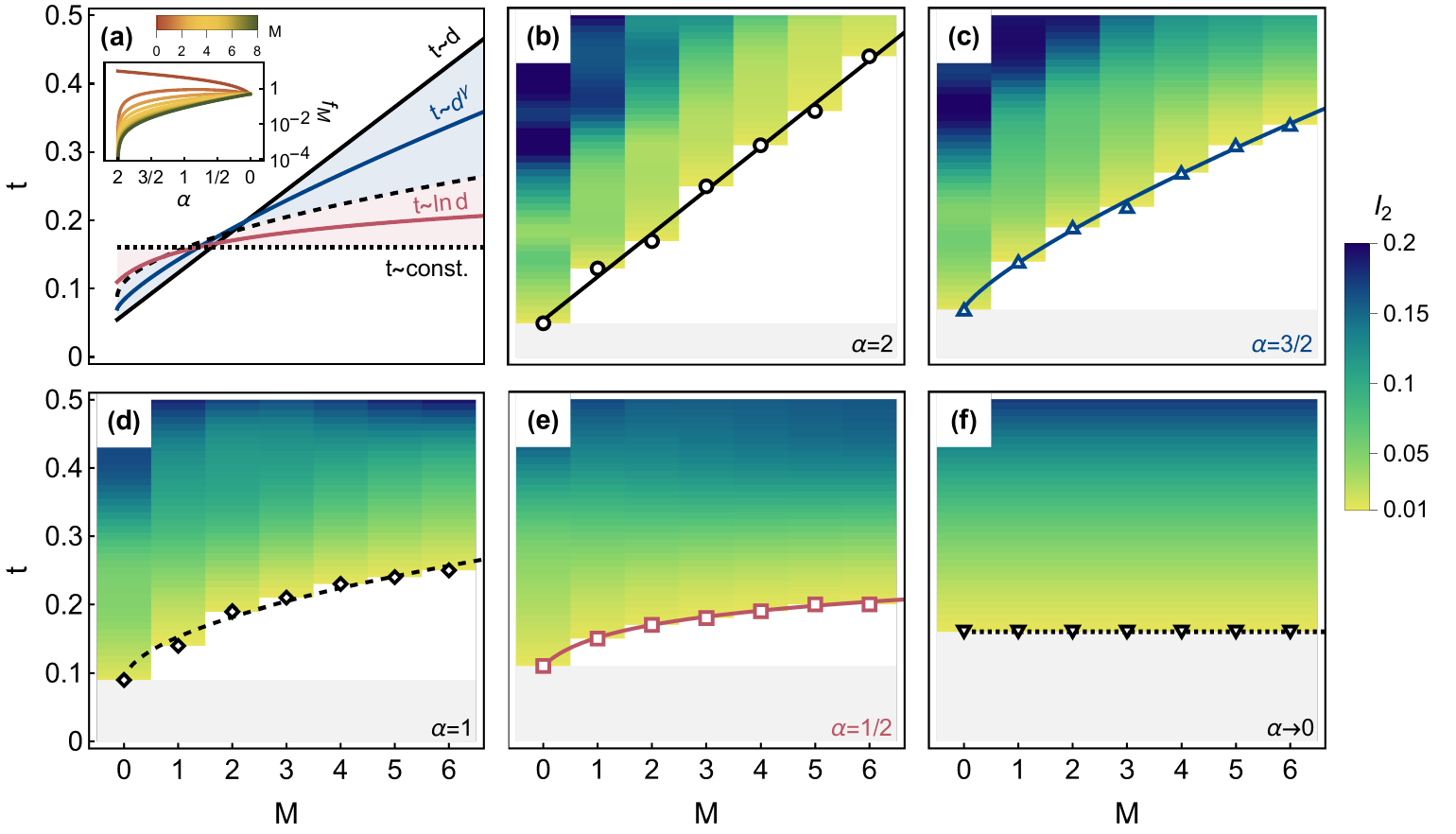}
    \caption{Spatiotemporal evolution of the Rényi-2 mutual information~\eqref{eq:Renyi2MI} of two modes ($l=\epsilon$) separated by a distance $d=(M+1) \epsilon$ after quenching the optical vacuum~\eqref{eq:PreQuenchHamiltonian} to the fractional Laplacian theory~\eqref{eq:FractionalLaplacianHamiltonian} with $L=2,m=1,\epsilon=0.1$ fixed. We vary the coupling range [see inset in \textbf{(a)} for the coupling strength $f_M = \abs{f_{j j+M} (\alpha)}$ of the two modes at various distances $M$] from $\alpha=2$ (relativistic) in \textbf{(b)} to $\alpha \to 0$ (all-to-all) in \textbf{(f)}. We cut off the mutual information below $0.01$ to probe the first correlation front (points) by fitting the characteristic curves summarized in Tab.~\ref{tab:LightCones} and shown in \textbf{(a)}. The light cones start to bend when $\alpha < 2$, as long-range couplings enable distant modes to become correlated at a faster rate. Correlations between neighboring modes pass the threshold at later times for longer range couplings, see gray regions in \textbf{(b)}--\textbf{(f)}, since the nearest-neighbor coupling strength $f_0$ decreases monotonically with $\alpha$, see inset in \textbf{(a)}, converging to $f_M \to 1/2$ when $\alpha \to 0$.}
    \label{fig:LightCones}
\end{figure*}

Finally, we investigate the rate at which two individual modes ($l=\epsilon$) at a distance $d = (M+1)\epsilon$ become correlated as the coupling range is varied. The onset of correlations is expected to follow (curved) light-cones as predicted by Lieb-Robinson bounds, which have been recently extended to long-range theories in the context of spin systems characterized by finite-dimensional Hilbert spaces~\cite{Tran2021} (see also~\cite{Chen2023}). Substantially fewer strict results are available for bosonic systems and the quasi-particle picture aside from numerical studies~\cite{Nezhadhaghighi2013, Nezhadhaghighi2014, Rajabpour2015}.

\renewcommand{\arraystretch}{1.2}
\begin{table}[b!]
    \centering
    \setlength{\tabcolsep}{8pt} 
    \begin{tabular}{r l l}
    \toprule
    Type & Form & Coupling range \\
    \midrule
    Linear & $t \gtrsim d$ & $\alpha = 2$ \\
    Algebraic & $t \gtrsim d^{\gamma}, \, \gamma \in (0,1)$ & $1 \le \alpha < 2$ \\
    Logarithmic & $t \gtrsim \ln d$ & $0 < \alpha < 1$ \\
    Constant & $t \gtrsim \text{const}.$ & $\alpha \to 0$ \\
    \bottomrule
    \end{tabular}
    \caption{Functional forms of the light cones describing the onset of correlations between two modes separated by a distance $d$ for varying coupling range $\alpha \in (0,2]$. The found regimes agree with those predicted in~\cite{Tran2021,Chen2023}.}
    \label{tab:LightCones}
\end{table}
\renewcommand{\arraystretch}{1}

By exploiting the OTA's configurability, we simulate the fractional Laplacian theory~\eqref{eq:FractionalLaplacianHamiltonian} with $\alpha \in (0,2]$ and probe the build-up of correlations in space \textit{and} time with the Rényi-2 mutual information following~\eqref{eq:Renyi2MI}, see Fig.~\hyperref[fig:LightCones]{5}, where $L=2,m=1,\epsilon=0.1$ such that $N=20$ as in the previous example. Depending on the range $\alpha$, we observe the light cones to follow characteristic curves, see Fig.~\hyperref[fig:LightCones]{5(a)} and Tab.~\ref{tab:LightCones} for an overview. 

As discussed in Sec.~\ref {subsec:Dynamical entanglement structure}, correlations follow straight light cones in a relativistic theory ($\alpha = 2$), with the incline given by the fastest quasi-particle velocity, see Fig.~\hyperref[fig:LightCones]{5(b)}. This is caused by the coupling strength $f_M = \abs{f_{j j+M} (\alpha)}$ between two modes separated by $M$ modes [see inset in Fig.~\hyperref[fig:LightCones]{5(a)}] being non-zero only if the modes are adjacent, i.e., $f_M \propto \delta_{M0}$ when $\alpha = 2$. Local correlations between neighboring modes require a finite time to reach the threshold $I_2 = 0.01$, as indicated by the gray region. 

In the regime $1 \le \alpha < 2$, see Figs.~\hyperref[fig:LightCones]{5(c)-(d)}, the light cones start to bend, demonstrating faster correlation propagation for large distances, as expected for long-range couplings characterized by $f_M > 0$ for all $M$. More precisely, the light cones follow algebraic curves. Simultaneously, nearest-neighbor correlations appear delayed, which is caused by a decreasing coupling strength between neighboring modes, as shown in the upper curve of the inset in Fig.~\hyperref[fig:LightCones]{5(a)}. Upon further increasing the coupling range to $0 < \alpha < 1$, the light cones become logarithmic, see Fig.~\hyperref[fig:LightCones]{5(e)}, until they disappear in the limit $\alpha \to 0$, see Fig.~\hyperref[fig:LightCones]{5(f)}. The latter case corresponds to an all-to-all coupling of the modes with $f_M \to 1/2$ for all distances $M$, allowing for immediate information transfer and rendering all modes \textit{pari passu}.

Remarkably, our findings are in perfect agreement with recent results for spin systems~\cite{Tran2021,Chen2023} at a moderate system size of $N=20$ modes. We stress that only \textit{two} of the twenty modes have to be measured and that the maximum required squeezing decreases approximately linearly with coupling range $\alpha$, with the maximum occurring at $\alpha = 2$ for which $\abs{z_{\text{max}}} \approx 0.35 \approx 3$dB. This highlights the prospects of the OTA for systematically engineering and studying the dynamics of quantum information in various field-theoretic settings beyond the simple case of a relativistic scalar.

\section{Error analysis}
\label{sec:Error analysis}
Photonic devices, like any experimental realization, have imperfections. Therefore, it is crucial to study the effects of experimental errors on our benchmarks. We focus on the two predominant error mechanisms in Boson-Sampler setups (see Fig.~\ref{fig:Boson sampler}): noise due to parameter fluctuations during state preparation and photon loss within the interferometer~\cite{Slussarenko2019}. We investigate the relativistic theory considered in Sec.~\ref{subsec:Dynamical entanglement structure}, see Fig.~\ref{fig:EntanglementGeneration}.

\subsection{State preparation noise}
\label{subsec:Noise}

The state entering the interferometer is a squeezed vacuum state with a phase. During its preparation, both the intensity $\xi$ and the phase $\Phi$ fluctuate from one experimental run to another. We remark that numerical simulations with tensor networks indicate that Boson Sampling remains classically hard in the presence of phase fluctuations~\cite{Paryzkova2025}; see also~\cite{Stefszky2025} for a recent sampling experiment discussing phase noise.

\begin{figure}[t!]
    \centering
    \includegraphics[width=0.99\columnwidth]{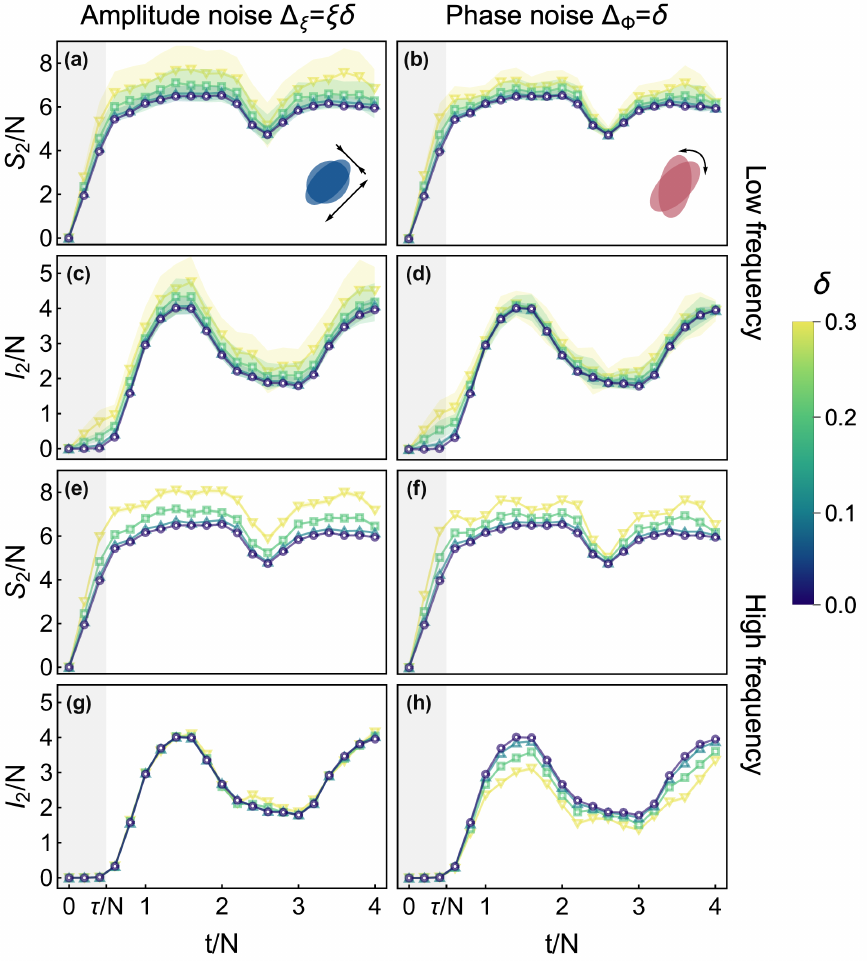}
    \caption{Effects of state preparation noise for the relativistic quench, cf. Sec.~\ref{subsec:Dynamical entanglement structure}. We consider the same setup underlying Fig.~\ref{fig:EntanglementGeneration}, that is, $N=25$ modes, a subinterval of length $\ell=L/5$ for the Rényi-2 entropy and two intervals of the same length separated by a distance $d=\ell$ for the Rényi-2 MI, as well as $m=1$, $\epsilon=2$. We sample $\mathcal{N}_{\text{S}}=100$ realizations of the normally-distributed optical parameters $\xi$ (squeezing amplitude, left column) and $\Phi$ (phase, right column) with fluctuations controlled via the noise parameter $\delta \in \{0, \, 0.1, \, 0.2, \, 0.3\}$. The shaded areas cover one standard deviation. For high-frequency noise, we resampled $100$ times to obtain the $\pm 1$\,s.d. interval, which turned out smaller than the line width. All information measures shift upwards for low-frequency noise, see panels \textbf{(a)}--\textbf{(d)}, while high-frequency noise mimics mixing, which increases local entropy and decreases correlations, as shown in panels \textbf{(e)}--\textbf{(h)}. All qualitative features of quench dynamics are preserved up to noise levels of $30\%$.}
    \label{fig:Noise}
\end{figure}

We simulate noise by randomly selecting the optical parameters from a Gaussian distribution centered around the ideal values. Figure~\ref{fig:Noise} shows the results obtained with $\mathcal{N}_{\text{S}}=100$ such noise realizations. The standard deviation $\Delta$ of the distribution determines the noise strength. As we expect fluctuations in the squeezing amplitude to increase with the amount of squeezing, we consider $\Delta_{\xi} = \xi \delta$ (left column). In contrast, we sample the phase from a Gaussian with constant width $\Delta_{\Phi} = \delta$ (right column). In both cases, $\delta > 0$ is a noise parameter with $\delta = 0$ corresponding to an ideal, noiseless input.

We distinguish two scenarios characterized by the time scales over which parameters fluctuate. If the input state can be prepared repeatedly with stable experimental parameters over the measurement time scale for a given set of model parameters, e.g., one time step of the evolution, we refer to the noise as low frequency, see Figs.~\hyperref[fig:Noise]{6(a)-(d)}. In this case, we directly compute information measures for each of the $\mathcal{N}_{\text{S}}$ randomly drawn parameters and average the result, e.g., $\langle S_{2}(\sigma) \rangle$. In a second scenario, circuit parameters fluctuate on time scales shorter than the measurement window, thereby affecting the estimation of individual covariance matrices and thus referred to as high-frequency, see Figs.~\hyperref[fig:Noise]{6(e)-(h)}. Here, we instead average the covariance $\sigma$ over $\mathcal{N}_{\text{S}}$ noise realizations before computing the quantities of our interest, e.g., $S_{2}(\langle\sigma\rangle)$. We note that low-frequency noise can always be converted to high-frequency noise by randomizing the ordering of covariance matrix measurements and model parameter choices.

For low-frequency noise, we find that both the entropy and the mutual information slightly increase with respect to the noiseless case. This can be understood by considering the relation between the number of quasi-particles present, $n$, and the optical parameters.
For the relativistic theory~\eqref{eq:Relativistic Real Lagrangian Density}, a momentum mode $k$ is populated by $n_k = \sinh^2 z_k$ quasi-particles, see Eq.~\eqref{eq: Number of quasi-particles}, discussed in App.~\ref{subsec:Quantum quench population}, with Eq.~\eqref{eq:OTAParameters2} and $\gamma_k = \epsilon \omega_k$ understood. Hence, symmetrically distributed deviations from the ideal values for $\xi$ and $\Phi$ are biased towards higher quasi-particle numbers, resulting in an overall growth of correlations.

For high-frequency noise, the observed increase in entropy and decrease in mutual information is explained by their elementary properties: mixing probability distributions increases their entropy and washes out correlations~\cite{Nielsen2010}. Remarkably, the mutual information remains almost unaffected by amplitude noise, see Figs.~\hyperref[fig:Noise]{6(g)}.
In fact, the covariance matrix of a (complex) squeezed vacuum state has diagonal elements $\cosh(2\xi) \mp \cos(\Phi) \sinh(2 \xi)$ and off-diagonal elements $- \sin(\Phi) \sinh(2 \xi)$. Hence, averaging over different amplitudes affects the covariance matrix elements in similar ways. When inserting the latter into the defining equation~\eqref{eq:Renyi2MI}, this causes cancellations of fluctuations to first order. In contrast, phase fluctuations do not correspond to global modifications.

\subsection{Photon Loss}
\label{subsec:Loss}

Losses are the main bottleneck of photonic platforms~\cite{Slussarenko2019}. Most photons are lost in the interferometer due to the many integrated beam splitters and phase shifts. Over the past decade, losses have posed significant challenges to the race toward quantum advantage, as sufficiently high loss levels can render boson sampling efficient on classical hardware~\cite{Bulmer2022,Oh2024a,Oh2024b,Cilluffo2025}.

\begin{figure}[t]
    \centering
    \includegraphics[width=0.99\columnwidth] 
    {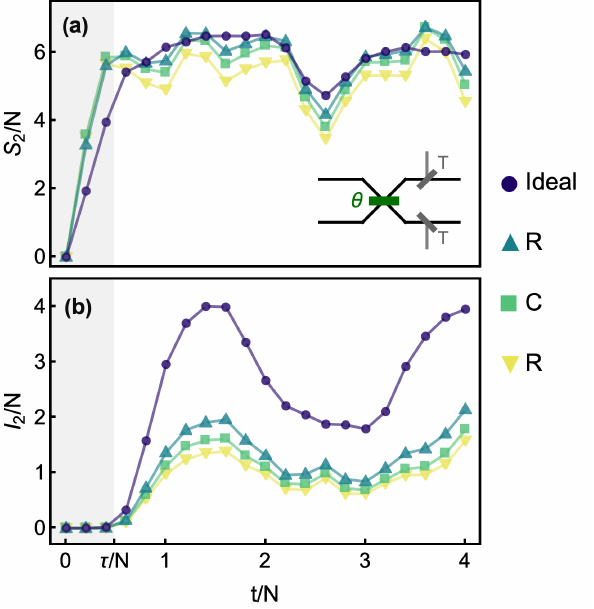}
    \caption{Effects of a lossy interferometer for the relativistic quench, cf. Sec.~\ref{subsec:Dynamical entanglement structure}. We consider the same setup underlying Fig.~\ref{fig:EntanglementGeneration}, that is, $N=25$ modes, a subinterval of length $\ell=L/5$ for the Rényi-2 entropy, see \textbf{(a)}, and two intervals of the same length separated by a distance $d=\ell$ for the Rényi-2 mutual information, see \textbf{(b)}, as well as $m=1$, $\epsilon=2$. We apply a loss channel to both output modes of each beam splitter by interfering with the vacuum at another beam splitter with transmittivity $T=0.98$ (see inset), resulting in $\sim 61.6\%$ overall transmittance. We compare three decomposition geometries (Reck up: blue triangles, Clements: green squares, Reck down: yellow triangles) with the ideal case (purple points). While the entropy remains robust to losses, having fewer quasi-particles reduces the mutual information. Nevertheless, qualitative features remain unaffected by losses.}
    \label{fig:Loss}
\end{figure}

The standard loss model in quantum optics couples each mode emerging from an (ideal) beam splitter to an ancillary vacuum mode via a beam splitter with transmissivity $T \in [0,1]$, which controls the loss rate ($T=1$ amounts to the lossless case). This is referred to as the symmetric loss model, see the inset in Fig.~\hyperref[fig:Loss]{7(a)} for a sketch. The corresponding quantum channel acts on every two-mode covariance matrix $\sigma$ as~\cite{Serafini2017}
\begin{equation}
    \varepsilon(\sigma) 
    = T S_{\text{BS}} \sigma S_{\text{BS}}^{\intercal}
    + \tfrac{1}{2}(1 - T)
    \mathds{1}_{4}.
\end{equation}
For a generic input covariance matrix $\sigma$, we compute $\varepsilon (\sigma)$ using a Python package available under~\cite{GitHubForLoss}.

An important factor influencing the impact of losses on the simulation is the interferometer geometry. Given the structural differences between the two predominant schemes, namely Reck's triangular~\cite{Reck1994} and Clement's~\cite{Clements2016} square meshes, we study both cases. For the former, we further distinguish between configurations where the beam splitters are arranged as a triangle pointing upwards and those where it points downwards (see Fig.~\ref{fig:OTAwithReck} in App.~\ref{subsubsec:BeamSplitterArray} for a pictorial representation of the down-oriented configuration), while the latter is homogeneous by construction. 

In Fig.~\ref{fig:Loss}, we show the results for the relativistic quench and the three interferometer configurations (Reck up: blue triangles, Clements: green squares, Reck down: yellow triangles) at $2\%$ losses per mode, which translates into an overall loss of $\sim 38.4\%$ for the Clements decomposition. This corresponds to typical loss rates of commercially available multi-mode interferometers~\cite{Stefszky2025}. For all three decompositions, the lossy data are in qualitative agreement with the ideal case (purple points). The Rényi entropy, see Fig.~\hyperref[fig:Loss]{7(a)}, remains robust to loss, which we attribute to two competing effects. On the one hand, loss decreases overall state purity and thus increases entropy. On the other hand, losses decrease the mean photon number, rendering the states more similar to the vacuum, and thus decreasing subsystem entropy. In contrast, losses generally reduce correlations, as seen in Fig.~\hyperref[fig:Loss]{7(b)}. 

We further observe a consistent ordering of information measures with respect to the decomposition scheme: both quantities of our interest are largest (smallest) for the upward (downward) Reck scheme, while Clements' scheme is positioned in between. This is because we consider the entropy of the 5-mode subsystem formed by the first five modes located at the upper end of the system. Hence, when the tip of the Reck triangle points upward, less noisy beam splitters are encountered. Instead, Clements' decomposition does not break spatial translation symmetry, and thus loss affects all modes equally. Homogeneous losses make loss effects easier to interpret and control, which is why Clements' decomposition is often preferred.

In summary, none of the considered experimental error sources alter the qualitative features of the simulated information dynamics, and error-induced deviations are moderate in experimentally realistic parameter regimes.

\section{Discussion}
\label{sec:Discussion}
We established the Optical Time Algorithm (OTA) for quantum-simulating the dynamics of various free scalar field theories. Our encoding is both intuitive and efficient: a single layer of phase shifters evolves the eigenmodes in time, while the physical properties of the underlying theory are encoded in two fixed interferometer arrays followed by single-mode squeezers. In this way, a wide landscape of QFT dynamics can be quantum-simulated by adapting the parameters of a few optical elements, including (real- and complex-valued) relativistic, non-local, and non-relativistic theories as well as curved spacetimes (also exotic combinations of the latter can be envisioned), with little to no constraints on the theories' parameters, the simulation time, or the spacetime dimension. The structure of the OTA overcomes the key constraint of other decompositions: the dynamics is entirely encoded in a single phase-shift layer, while other components remain time-independent. 

Regarding applications, we focused on the light-cone-like spreading of quantum entanglement and correlations after a quench for varying coupling range. We considered initial product coherent states in which case the OTA reduces to a squeezing layer followed by passive linear optical elements, resembling the setup of boson sampling experiments. We found that $10$ to $20$ modes are sufficient to capture field-theoretic predictions, also under realistic experimental constraints, while only a handful of those modes need to be measured. The tight link between the OTA and Gaussian boson sampling highlights the OTA's prospects for addressing classically hard problems on quantum hardware.

We envision four main routes for generalizations and further applications. First, in analyzing the information dynamics of quantum fields, we have focused on pure Gaussian initial states such as the vacuum, in which case the OTA attains an elementary form. This was mainly motivated by the ubiquity of Gaussian states in quantum optics experiments. However, the OTA does not rely on simple input states. Given the growing range of techniques for preparing non-Gaussian states in the lab, for instance, via conditioning schemes~\cite{Walschaers2021,Chabaud2023,Stefszky2025}, the OTA is readily applicable for studying non-Gaussian scenarios. Given the inherent complexity of non-Gaussian states~\cite{Mele2025}, this approach may offer advantages over classical methods even when relying on homodyne measurements. In this context, it is also of interest to consider conditioning the particle numbers at only some of the output modes using photon-number-resolving detectors, for which the experimental setup is precisely given by the Gaussian boson sampler. This would grant access to the symmetry sectors of the local particle number operator, which, given a suitable readout technique, could enable the readout of symmetry-resolved entanglement entropies in close-to-continuum settings~\cite{Lukin2019,Goldstein2018}.

Second, while the OTA maintains its form in time-dependent scenarios whenever the diagonalization transformation (realized by the interferometers) remains time-independent, the situation becomes more complex when considering general time dependencies, as is evident, for example, in driven Bose-Einstein condensates~\cite{Bluvstein2021}. In this case, acting with the unitary time-evolution operator on the quadratures requires including time-ordering in the exponential of the corresponding symplectic matrix transformation. This induces time dependencies in the symplectic eigenvalues, thereby complicating the disentangling of temporal contributions.

Third, since the OTA encompasses all relevant free theories, a natural extension is to incorporate interactions~\cite{Hartmann2016}. The experimental realization of strong interactions at the level of a few photons is challenging, as nonlinear optical effects in conventional materials are typically negligible at the associated light powers. In the last two decades, significant advances have been made towards this goal~\cite{Chang2014}, e.g., by interfacing optical modes with single atoms or atomic ensembles in cavity QED systems~\cite{Walther2006}, by using hybrid states of light and atomic media~\cite{Bajcsy2009, Peyronel2012,Firstenberg2016}, or by resorting to quantum dots~\cite{Nielsen2025}. However, none of the present approaches has been demonstrated to be scalable. We speculate that the key idea of our approach, that is, isolating the time-dependence and encoding the Hamiltonian into a fixed optical circuit, is generalizable to weakly interacting theories in a perturbative fashion. Beyond the perturbative regime, a Trotterization of the dynamics might be necessary, in which case the OTA would form the elementary building block of the non-interacting part. 

Fourth, recent advances towards scalable multimode photonic platforms using spatial~\cite{Wang2020,Zhong2020, Brandt2020,Goel2024,Wang2024}, temporal~\cite{Madsen2022,Bouchard2024}, and frequency~\cite{Folge2024} degrees of freedom of light provide exciting opportunities for experimental realizations of the quantum field simulations proposed in this work. Although we emphasized photonic implementations due to their high configurability, other setups involving bosonic modes can also be envisioned, including circuit QED~\cite{Blais2021} and ultracold atoms~\cite{Pitaevskii2016} trapped in highly configurable optical potentials~\cite{Gross2017,Schaefer2020}.

\section*{Acknowledgements}
We thank Benjamin Brecht, Dario Cilluffo, Martin Plenio, Christine Silberhorn, René Sondenheimer, and Michael Stefzsky for useful discussions. We thank Alexander Naumann and Robin Strahlendorf for providing a Python package for simulating photon loss~\cite{GitHubForLoss}. T.H. acknowledges support from the European Union under project ShoQC within the ERA-NET Cofund in Quantum Technologies (QuantERA) program, the F.R.S.-FNRS under project CHEQS within the Excellence of Science (EOS) program, the BMBF project PhoQuant (Grant No. 13N16110), and the EU project SPINUS (Grant No. 101135699). M.D. and M.G. are supported by funding from the German Research Foundation (DFG) under the project identifier 398816777-SFB 1375 (NOA).

\section*{Data availability and source code}
We provide the code used for generating Figs.~\ref{fig:EntanglementGeneration} --~\ref{fig:Loss} and~\ref{fig:EntanglementGenerationOBC} together with an algorithm that computes all circuit parameters given a Hamiltonian matrix $H$ via the GitHub repository~\cite{GitHub}. The code for loss simulation is publicly accessible via GitHub~\cite{GitHubForLoss}.


\appendix

\section{Proofs}
\label{sec:Proof}

\subsection{Proof of Eq.~\eqref{eq:IsolatingTime}}
\label{subsec:Proof1}
The isolation of the time parameter into single-mode phase shifters is achieved by generalizing Williamson's theorem for the Hamiltonian matrix $H = S_{\text{SD}}^{\intercal} D S_{\text{SD}}$ [see Eq. \eqref{eq:WilliamsonsTheorem}] to the symplectic matrix $S(t)$ describing time evolution. After introducing the corresponding unitary transformation $\boldsymbol{U}_{\text{SD}}$ defined via $S_{\text{SD}} \boldsymbol{r} = \boldsymbol{U}_{\text{SD}}^{\dagger} \boldsymbol{r} \boldsymbol{U}_{\text{SD}}$, the Hamiltonian becomes
\begin{equation}
    \boldsymbol{H}
    = \dfrac{1}{2} \boldsymbol{r}^{\intercal}
    H \boldsymbol{r}
    = \dfrac{1}{2}
    \boldsymbol{U}_{\text{SD}}^{\dagger}
    \boldsymbol{r}^{\intercal}
    ( 
    D_{N} \oplus D_{N}
    )
    \boldsymbol{r}
    \boldsymbol{U}_{\text{SD}} .
    \label{eq:HamiltonianSymplecticDiagonalization}
\end{equation}
By linearity, the corresponding unitary can be rewritten as a product of time-dependent phase shifts between two time-independent unitaries [cf. Eq.~\eqref{eq:PhaseShift}]
\begin{equation} \label{eq:Williamson's theorem applied to S}
    \boldsymbol{U}(t)
    = e^{-it \boldsymbol{H}}
    = \boldsymbol{U}_{\text{SD}}^{\dagger}
    \boldsymbol{U}_{\text{PS}}(\varphi(t))
    \boldsymbol{U}_{\text{SD}} ,
\end{equation}
where $\varphi(t) = t (d_{1}, \ldots, d_{N})^{\intercal}$. At the level of symplectic matrices, the previous decomposition reads
\begin{equation} \label{eq:Williamson's theorem applied to S}
    S(t) 
    = S_{\text{SD}}^{-1}
    S_{\text{PS}}(\varphi(t))
    S_{\text{SD}}. 
\end{equation}

\subsection{Proof of the OTA}
\label{subsec:Proof2}

\subsubsection{Eigensystems of $H^{\phi}$ and $H^{\pi}$}
\label{subsubsec:Eigensystem}
For the OTA, we consider the class of block diagonal Hamiltonians with commuting blocks $H^{\phi}$ and $H^{\pi}$. Given Eq.~\eqref{eq:Williamson's theorem applied to S}, our goal is to further decompose the SD matrix by means of optical transformations. We make the ansatz
\begin{equation}
    S_{\text{SD}} = \begin{pmatrix}
        \mathds{O}_{N} & A_{N} \\
        B_{N} & \mathds{O}_{N}
    \end{pmatrix},
    \label{eq:SSDAnsatz}
\end{equation}
with $A_N$ and $B_N$ being further constrained as $S_{\text{SD}}$ must diagonalize the Hamiltonian matrix $H$ and has to be symplectic. When inserting \eqref{eq:SSDAnsatz} into \eqref{eq:WilliamsonsTheorem}, the former condition translates into the two constraints
\begin{equation} \label{eq:SD constraints 1}
    A_{N}^{\intercal} D_{N} A_{N} = H^{\pi}, \quad B_{N}^{\intercal} D_{N} B_{N} = H^{\phi}.
\end{equation}
The latter condition can be evaluated by demanding that \eqref{eq:SSDAnsatz} preserves the symplectic form $\Omega$, which leads to
\begin{equation}
    A_{N}^{\intercal} B_{N} = - \mathds{1}_{N}.
    \label{eq:SD constraints 2}
\end{equation}

The constraints ensure $B^{\intercal}_{N}$ to be the (real) eigenvector matrix diagonalizing both $H^{\phi}$ \textit{and} $H^{\pi}$. Then, multiplying the first two constraints and using the third to remove $A_{N}$, results in
\begin{equation} \label{eq:Product Similarity}
    D^{2}_{N}
    = (B_{N}^{\intercal})^{-1}
    H^{\phi} H^{\pi}
    B_{N}^{\intercal} ,
\end{equation}
which implies Eq.~\eqref{eq:Eigenvalues product}.
Further, the constraints fix the normalization of $B_{N} B_{N}^{\intercal}$: isolating $B_{N} B_{N}^{\intercal} = B_{N} H^{\phi} B_{N}^{-1} D_{N}^{-1}$ from the second constraint and using the similarity relation for $H^{\phi}$ on the right hand side, which itself can be rewritten in terms of $B_{N}^{-1}$ since $H^{\phi}$ is symmetric, leads to
\begin{equation} \label{eq:BB=Gamma}
    B_{N} B_{N}^{\intercal}
    = \text{diag}(\lambda^{\phi}_{1}/d_{1}, \dots, \lambda^{\phi}_{N}/d_{N}) 
    \equiv \Gamma_{N}.
\end{equation}
By using Eq.~\eqref{eq:Eigenvalues product}, it becomes evident that
the elements of the diagonal matrix $\Gamma_{N} = \text{diag}(\gamma_1, \dots, \gamma_N)$ are related to the eigenvalues of $H^{\phi}$ and $H^{\pi}$ via Eq.~\eqref{eq:Eigenvalues ratio}. We remark that the same line of reasoning applies when working with $A_N$ instead of $B_N$.

\subsubsection{Squeezer layer and gauge fixing}
\label{subsubsec:SqueezerLayer}
Since the normal decomposition is not unique~\cite{deGosson2006,Son2021}, the SD matrix \eqref{eq:Williamson's theorem applied to S} is defined only up to an orthosmyplectic transformation\footnote{At the level of $H$, this is equivalent to $G^{\intercal} D G = D$.} $S_{\text{SD}} \to G S_{\text{SD}}$, with $G \in OrSp(2N)$ leaving the phase shifter array invariant, i.e.,
\begin{equation} \label{eq:Gauge transformation}
    G^{-1} S_{\text{PS}} (\varphi(t)) G = S_{\text{PS}} (\varphi(t)) .
\end{equation}
We refer to this mathematical redundancy encapsulated by $G$ as a \textit{gauge transformation}. 

The freedom of choosing the gauge comes in handy when decomposing $S_{\text{SD}}$ into optical elements, for which we employ the (symplectic) polar decomposition~\cite{Houde2024}
\begin{equation}
    S_{\text{SD}} = \Sigma P.
    \label{eq:PolarDecomposition}
\end{equation}
Here, we introduced the symplectic, symmetric, and positive semi-definite matrix $\Sigma = (S_{\text{SD}} S_{\text{SD}}^{\intercal})^{1/2}$, while $P = \Sigma^{-1} S_{\text{SD}}$ is orthosymplectic. After evaluating the matrix $\Sigma^2$ for \eqref{eq:SSDAnsatz} in some general gauge $G$, we find
\begin{equation}
    \Sigma^2 = \begin{pmatrix}
        G_2 \Gamma_N G_2^{\intercal} + G_1 \Gamma^{-1}_N G_1^{\intercal} & G_2 \Gamma_N G_1^{\intercal} - G_1 \Gamma^{-1}_N G_2^{\intercal} \\
        G_1 \Gamma_N G_2^{\intercal} - G_2 \Gamma^{-1}_N G_1^{\intercal} & G_1 \Gamma_N G_1^{\intercal} + G_2 \Gamma^{-1}_N G_2^{\intercal}
    \end{pmatrix},
\end{equation}
where we used \eqref{eq:SD constraints 2}. 

The latter matrix decouples when choosing the gauge $G_{1} = \mathds{O}_{N}$ and $G_{2} = \mathds{1}_{N}$, in which case
\begin{equation}
    \Sigma = \Gamma_N^{1/2} \oplus \Gamma_N^{-1/2} = S_{\text{Sq}} (z),
    \label{eq:SigmaSqueezer}
\end{equation}
represents a squeezer with $z_j = - (1/2) \ln \gamma_j$ [cf. Eq.~\eqref{eq:Squeezer}], as desired. Further, the remaining matrix $P$ indeed encodes the orthonormal eigenvectors of $H^{\phi}$ (or $H^{\pi}$) since
\begin{equation}
    P = (\Gamma_N^{-1/2} B_N) \oplus (\Gamma_N^{-1/2} B_N) \equiv P_N \oplus P_N,
\end{equation}
where we recall that $\Gamma_N$ is diagonal. Note also that the symplectic constraint~\eqref{eq:SD constraints 2} becomes $B_N = - \Gamma_N A_N$, confirming that $A_N$ and $B_N$ are proportional.

We remark that our choice of gauge fixing improves the efficiency over a direct implementation of $S_{\text{SD}}$ as in \eqref{eq:SSDAnsatz}: choosing $G_1 = \mathds{1}_N$ and $G_2 = \mathds{O}_N$ (corresponding to $G=\mathds{1}_{2N}$) leads to $P_N$ appearing on the \textit{anti}-diagonal blocks of $P$ instead. This would require an additional layer of phase shifters---which amounts to rotating $P$ back to diagonal form---before the interferometer array.

\subsubsection{Decomposing the interferometer}
\label{subsubsec:BeamSplitterArray}

\begin{figure}[t!]
\centering
    \includegraphics[clip,width=0.85\columnwidth]{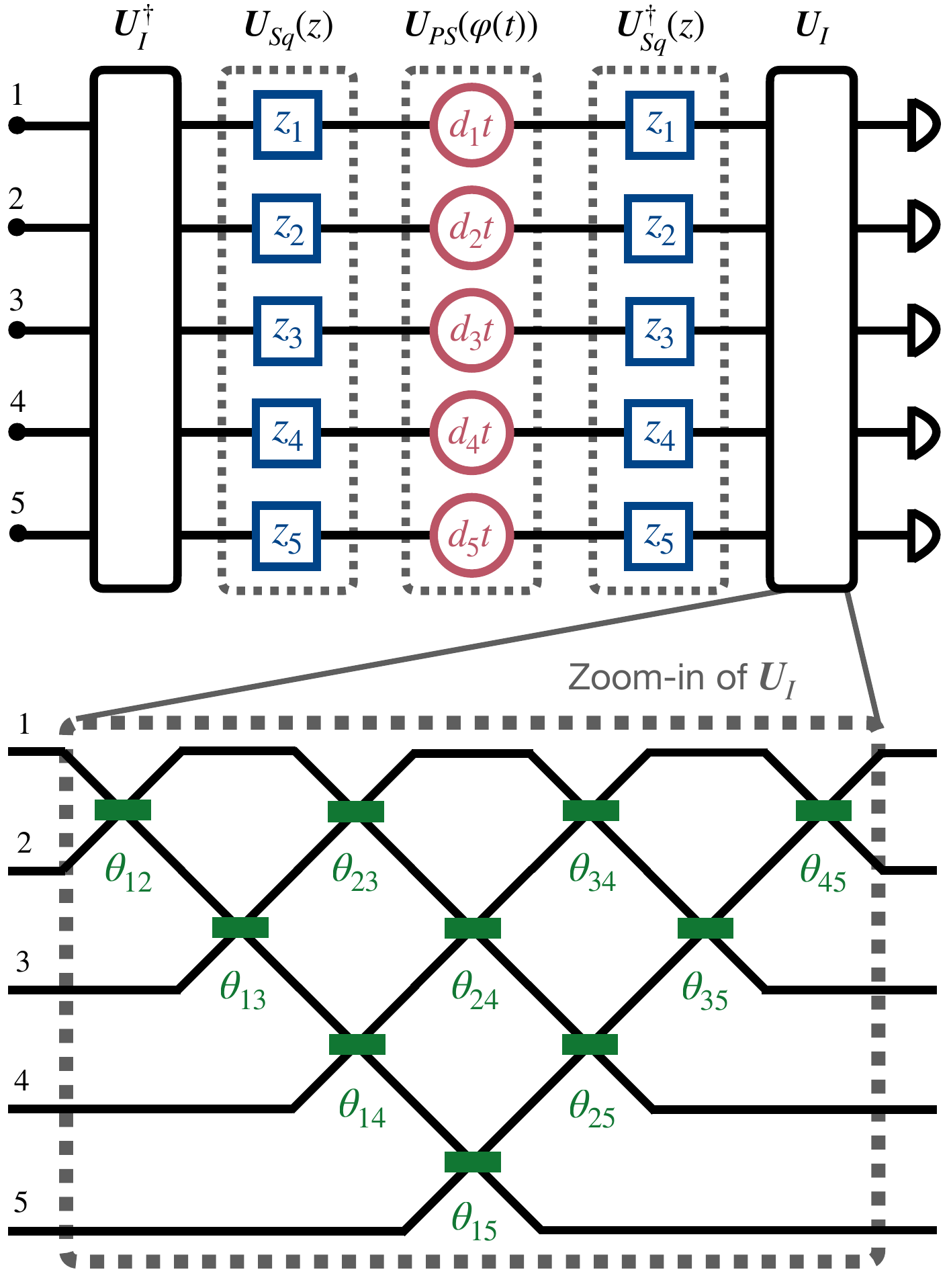}
    \caption{Pictorial representation of the OTA for $N=5$. The interferometer layers are decomposed here according to a Reck-inspired method.} 
    \label{fig:OTAwithReck}
\end{figure}

A variety of schemes can be implemented to decompose the interferometers, e.g. Reck's~\cite{Reck1994} and Clements'~\cite{Clements2016}, see~\cite{Cilluffo2024} for an overview. Inspired by the first method, as visible in the zoom-in in Fig.~\ref{fig:OTAwithReck}, we further decompose the orthosymplectic matrix $P_N$ by means of a QR decomposition
\begin{equation}
    P_N O_N = \mathcal{P}_N,
\end{equation}
where $\mathcal{P}_N = \text{diag} (\pm 1, \dots, \pm 1)$ and $O_N$ being an orthogonal matrix, which can be written as a matrix product over Givens rotations $O_{jk}(\theta)$ defined as
\begin{equation}
    O_{jk}(\theta) = \begin{pmatrix}
        1 & \cdots & 0 & \ldots & 0 & \cdots & 0 \\
        \vdots & \ddots & \vdots & & \vdots & & \vdots \\
        0 & \cdots & \cos(\theta) & \cdots &\sin(\theta) & \cdots & 0 \\
        \vdots &  & \vdots & \ddots & \vdots & & \vdots \\
        0 & \cdots & -\sin(\theta) & \cdots & \cos(\theta) & \cdots & 0 \\
        \vdots &  & \vdots &  & \vdots & \ddots & \vdots \\
        0 & \cdots & 0 & \ldots & 0 & \cdots & 1 \\
    \end{pmatrix}.
    \label{eq:GivensRotation}
\end{equation}
The latter contains the beam splitter transformation $S_{\text{BS}}(\theta_{jk})$, i.e., a rotation, at the $j$th and $k$th intersections between rows and columns. 

We engineer a sequence of Givens rotations that results in a diagonal matrix where all but the last entry are one, i.e., $\mathcal{P}_N^{\pm} = \text{diag}(1, \dots, \pm 1)$. Denoting by $p_{j k}$ the elements of the initial matrix $P_N$, we choose for all $k$ such that $1 \le k < N$, as long as there is a column $j > k$ with $p_{k j}$ different from zero, the beam splitter transmissivity $\theta$ and some $r>0$ such that
\begin{equation}
    r \cos \theta = p_{k k}, \quad - r \sin \theta = p_{k j} .
\end{equation}
When applying the Givens rotation implicitly defined via the latter equations to $P_N$ from the right $P_N \to P_N O_{jk} (\theta)$, the corresponding elements of the transformed matrix become $p_{k k} = r$ and $p_{k j} = 0$. Repeating this procedure for all $1 \le k < N$ and $j>k$ indeed results in
\begin{equation}
    P_{N} O_{1 2} O_{1 3} \dots O_{1 N} O_{2 3} \dots O_{N-1 N} = \mathcal{P}^{\pm}_N,
\end{equation}
which is equivalent to
\begin{equation}
    P = (\mathcal{P}_N^{\pm} \oplus \mathcal{P}_N^{\pm}) S^{-1}_{\text{BS}} (\theta),
    \label{eq:TBSRelation}
\end{equation}
with the triangular ordering, compatible with Fig.~\ref{fig:OTAwithReck},
\begin{equation}
    \begin{split}
        S_{\text{I}}
        & \equiv S_{\text{BS}} (\theta_{1 2}) S_{\text{BS}} (\theta_{1 3}) \dots S_{\text{BS}} (\theta_{1 N}) \\
        &\hspace{1.38cm}\times S_{\text{BS}} (\theta_{2 3}) \dots S_{\text{BS}} (\theta_{2 N}) \\
        &\hspace{2.15cm} \times \dots \\
        &\hspace{2,78cm} \times S_{\text{BS}}(\theta_{N-1 N}) ,
    \end{split}
    \label{eq:NModeBeamSplitter}
\end{equation}
understood. In this way, all beam splitter parameters can be identified iteratively.

The last step is to treat the two possible signs of $\mathcal{P}_N^{\pm}$ in Eq.~\eqref{eq:TBSRelation}. When the last entry is positive, we obtain $P = S^{-1}_{\text{I}}$, which is what we wanted to show. For the negative sign, we require an additional phase shift on the last mode $P = S_{\text{PS}} (0,\dots,\pi) S^{-1}_{\text{I}} $. However, this phase can be absorbed by adjusting the gauge: using that $S_{\text{PS}} (0) = S_{\text{PS}} (\pi) = - \mathds{1}_2$ is proportional to the identity, we can rewrite the full symplectic diagonalizing matrix as $S_{\text{SD}} = S_{\text{Sq}} (z) P =  S_{\text{PS}} (0,\dots,\pi) S_{\text{Sq}} (z) S^{-1}_{\text{I}}$, where we used the standard gauge fixing ($G_1 = \mathcal{O}_N, G_2 = \mathds{1}_N$). Transforming the gauge to $G \to S_{\text{PS}}^{-1} (0,\dots,\pi) G$, such that $G_2 = \mathcal{P}_N^{-}$, supersedes the additional phase shift, while leaving the squeezer unchanged. It is straightforward to check that this gauge still satisfies~\eqref{eq:Gauge transformation}, which completes our proof.

\subsection{Time-dependent case}
\label{subsubsec:Proof for time-dependent OTA}
We generalize the OTA to time-dependent Hamiltonian matrices $H = H(t)$ for time-independent eigenvectors $P \neq P(t)$, which ensures $[\boldsymbol{H} (t_1), \boldsymbol{H}(t_2)] = 0$. Starting once again from Williamson's theorem~\eqref{eq:WilliamsonsTheorem}, i.e., $H(t)=S_{\text{SD}}^{\intercal} (t) D(t) S_{\text{SD}} (t)$, and using the polar decomposition $S_{\text{SD}} = S_{\text{Sq}} (t) P$ [see Eqs.~\eqref{eq:PolarDecomposition} and~\eqref{eq:SigmaSqueezer}] results in 
\begin{equation}
    \boldsymbol{H}(t)
    = \dfrac{1}{2}
    \boldsymbol{U}_{\text{I}}
    \boldsymbol{r}^{\intercal}
    S_{\text{Sq}}^{\intercal} (t)
    D (t)
    S_{\text{Sq}} (t)
    \boldsymbol{r}
    \boldsymbol{U}_{\text{I}}^{\dagger},
\end{equation}
with $P = S_{\text{I}}^{-1}$ and $S_{\text{I}} \boldsymbol{r} = \boldsymbol{U}_{\text{I}}^{\dagger} \boldsymbol{r} \boldsymbol{U}_{\text{I}}$ understood. The corresponding unitary time evolution operator becomes
\begin{equation}
    \begin{split}
        \boldsymbol{U}(t) &= \exp \left[ - i \int_{0}^{t} \mathrm{d}t' \boldsymbol{H}(t') \right] \\
        &= \boldsymbol{U}_{\text{I}} \exp \left[  - \frac{i}{2} \boldsymbol{r}^{\intercal} \int_{0}^{t} \mathrm{d}t' S_{\text{Sq}}^{\intercal} (t') D (t') S_{\text{Sq}} (t') \boldsymbol{r} \right]  \boldsymbol{U}_{\text{I}}^{\dagger}.
    \end{split}
\end{equation}
As $\boldsymbol{U}_{\text{I}}$ diagonalizes the unitary, the remaining integral runs over diagonal matrices only. Hence, we recover the OTA when introducing time-integrated eigenvalues
\begin{equation}
    \tilde{\lambda}_{j}^{\phi(\pi)}
    = \int_{0}^{t} \mathrm{d}t' \lambda_{j}^{\phi(\pi)}(t'),
\end{equation}
which leads to
\begin{equation}
    S(t) = S_{\text{I}} \, S_{\text{Sq}}^{-1}[z(t)] \, S_{\text{PS}}[\varphi(t)] \, S_{\text{Sq}}[z(t)] \, S_{\text{I}}^{-1}.
\end{equation}
The single-mode optical parameters become time-dependent, i.e., $\varphi_j (t) = \tilde{d}_j (t), z_j (t) = -(1/2) \ln \tilde{\gamma}_j (t)$, with Eqs.~\eqref{eq:Eigenvalues ratio} and Eqs.~\eqref{eq:Eigenvalues product} generalizing to $\tilde{\gamma}_{j}^{2}(t) = \tilde\lambda^{\phi}_{j}(t)/\tilde\lambda^{\pi}_{j} (t)$ and $\tilde{d}_{j}^{2}(t) = \tilde\lambda^{\phi}_{j}(t)\tilde\lambda^{\pi}_{j}(t)$, respectively.

\subsection{Coherent inputs}
\label{subsec:CSq OTA Appendix}
Here, we show that the OTA dramatically simplifies (as discussed in Sec.~\ref{subsubsec:Simplification for vacuum inputs}) in the more general setting of coherent inputs by proving~\eqref{eq:Complex Squeezer Parameters} and a generalization of~\eqref{eq:Boson sampler simplification}. We define coherent states as displaced vacuum states $\ket{\alpha} = \boldsymbol{D}(\alpha) \ket{0}$, where $\boldsymbol{D}(\alpha) = e^{\alpha \boldsymbol{a}^{\dagger} - \alpha^{*} \boldsymbol{a}}$ with $\boldsymbol{a} = (\boldsymbol{a}_1, \dots, \boldsymbol{a}_N)^{\intercal}$ (analogously for the vector of complex phases $\alpha \in \mathbb{C}^N$). We recall that a single-mode coherent state $\ket{\alpha}$ is completely characterized by its complex amplitude $r_{\alpha} = \braket{\alpha | \boldsymbol{r} | \alpha}$ and the vacuum covariance $\sigma_{\alpha} = (1/2) \mathds{1}$, and that the dynamics is captured by $r (t) = S (t) r_{\alpha}$ and $\sigma (t) = S (t) S^{\intercal} (t)/2$. As passive transformations map coherent states onto coherent states~\cite{Weedbrook2012}, the first interferometer layer merely modifies the coherent states' amplitudes $\alpha \to \alpha'$ via $r_{\alpha'} = S_{\text{I}}^{-1} r_{\alpha}$, with the vacuum corresponding the special case $0 \to 0$. Hence, one may equally input $\ket{\alpha'}$ into the circuit and remove the first interferometer.

The main simplification for coherent inputs hinges on the fact that the single-mode (SM) part of the OTA, i.e.,
\begin{equation}
    S_{\text{SM}}(z, \varphi)
    = S_{\text{Sq}}^{-1}(z)
    S_{\text{PS}}(\varphi)
    S_{\text{Sq}}(z),
\end{equation}
can be rewritten as a complex squeezer, see Eq.~\eqref{eq:ComplexSqueezer}. To show this, we must enforce consistency of the quantum state's first and second moments only at the single-mode level. The latter implies the condition
\begin{equation}
    (S_{\text{SM}} S_{\text{SM}}^{\intercal}) (z,\varphi)
    = (S_{\text{CSq}} S_{\text{CSq}}^{\intercal})
    (\zeta),
\end{equation}
with $\zeta = \xi e^{i \Phi}$ understood, which is equivalent to the non-linear system of equations
\begin{equation}
    \begin{split}
        \cosh(2\xi) &= \cos^{2} (\varphi) + \sin^{2} (\varphi) \cosh(4z), \\ 
        \sin^{2} (\varphi) \sinh(4z) &= - \cos(\Phi) \sinh(2 \xi), \\
        \sin(2 \varphi) \sinh(2z) &= - \sin(\Phi) \sinh(2 \xi).
    \end{split}
\end{equation}
While the first equation amounts to the first equation in~\eqref{eq:Complex Squeezer Parameters}, dividing the third by the second gives
\begin{equation}
    \tan \Phi = \dfrac{\cot \varphi}{\cosh(2z)},
\end{equation}
thus proving the second equation in~\eqref{eq:Complex Squeezer Parameters} as well. This also completes the proof of~\eqref{eq:Boson sampler simplification} for vacuum inputs, as the first-moment condition turns trivial.

For coherent inputs, it remains to identify the mapping of the coherent amplitudes. From a geometrical perspective in phase space, we need to displace the modified initial state with amplitude $\alpha'$ yet another time when matching $S_{\text{SM}}$ with a complex squeezer. To show this, we switch to unitary operators and employ several commutation relations. First, we note that
\begin{equation}
    \begin{split}
        \boldsymbol{U}_{\text{SM}}(z, \varphi) \ket{\alpha'}
        &= \boldsymbol{U}_{\text{SM}}(z, \varphi) \boldsymbol{D}(\alpha') \ket{0} \\
        &= \boldsymbol{D}(\alpha'')
        \boldsymbol{U}_{\text{SM}}(z, \varphi) \ket{0},
    \end{split}
\end{equation}
where we used the elementary commutation relations
\begin{equation}
    \begin{split}
        \boldsymbol{U}_{\text{Sq}} (z) \boldsymbol{D} (\alpha') &= \boldsymbol{D} (\alpha' \cosh z - \alpha'^* \sinh z) \boldsymbol{U}_{\text{Sq}} (z), \\
        \boldsymbol{U}_{\text{PS}}(\varphi) \boldsymbol{D}(\alpha') &= \boldsymbol{D}(\alpha' e^{- i \varphi}) \boldsymbol{U}_{\text{PS}}(\varphi),
    \end{split}
\end{equation}
leading to $\alpha''$ depending on $\alpha', z$ and $\varphi$ through
\begin{equation}
    \alpha''= \alpha' \cos \varphi - i \sin \varphi [\alpha' \cosh(2z) - \alpha'^{*} \sinh(2z)].
\end{equation}
As we have already established $\boldsymbol{U}_{\text{SM}}(z, \varphi) \ket{0} = \boldsymbol{U}_{\text{CSq}} (\zeta) \ket{0}$, the last step is to commute $\boldsymbol{D}(\alpha'') \boldsymbol{U}_{\text{CSq}}(\zeta) = \boldsymbol{U}_{\text{CSq}}(\zeta) \boldsymbol{D}(\alpha''')$, which requires
\begin{equation}
    \alpha''' = \alpha'' \cosh\xi + \alpha''^{*} e^{i\Phi} \sinh \xi,
\end{equation}
thereby completing the proof.

\section{Circuit parameters for the relativistic scalar on $N=5$ modes}
\label{sec:Complete decomposition N=5}
To illustrate the decomposition scheme, we provide all circuit parameters for the relativistic theory~\eqref{eq:Relativistic Real Hamiltonian Operator in Real Space} when $N=5$. The phases $d_{j}$ reduce to the relativistic dispersion~\eqref{eq:Dispersion relation real relativistic theory}, while the squeezing parameters are $z_j = -(1/2)\ln (\epsilon d_j)$.
The beam splitter parameters $\theta_{ij}$ follow after a straightforward numeric computation using the iterative scheme described in App.~\ref{subsubsec:BeamSplitterArray}. We report all the parameters for $m=1$ and $\epsilon=2$ in Tab.~\ref{tab:ParametersN5}.

\renewcommand{\arraystretch}{1.2}
\begin{table}[h!]
    \centering
    \setlength{\tabcolsep}{4pt} 
    \begin{tabular}{c | c c c c c}
    \toprule
    $\theta_{i j}$ & $j=1$ & $2$ & $3$ & $4$ & $5$ \\
    \midrule
    $i=1$ & & $2.36$ & $0.62$ & $0.52$ & $0.46$ \\
    $2$ & & & $0.52$ & $-0.62$ & $-0.79$ \\
    $3$ & & & & $-1.02$ & $0$ \\
    $4$ & & & & & $-1.57$ \\
    $5$ & & & & & \\
    \midrule
    $d_{j}$ & $1$ & $1.16$ & $1.38$ & $1.38$ & $1.16$ \\
    \midrule
    $z_{j}$ & $-0.35$ & $-0.42$ & $-0.51$ & $-0.51$ & $-0.42$ \\
    \bottomrule
    \end{tabular}
    \caption{Optical parameters for the relativistic theory on $N=5$ modes with $m=1$ and $\epsilon=2$.}
    \label{tab:ParametersN5}
\end{table}

\section{Fractional Laplacian theory on a finite lattice}
\label{sec:Fractional Laplacian Theory in Momentum Space}
In flat space, the Laplace operator $-\partial_x^2$ is diagonalized by a Fourier transform to momentum space. Hence, arbitrary powers of the Laplace operator $(-\partial_x^2)^{\alpha/2}$ can be treated by raising its momentum-space eigenvalues to the same power and then transforming back to position space~\cite{Lischke2020}. The eigenvalue of the term $\phi (-\partial^2_{x})^{\alpha/2} \phi$ is dictated by the fractional dispersion relation~\eqref{eq:FractionalDispersion} in the massless limit $m \to 0$. Adapting this logic to the periodic lattice results in
\begin{equation}
    f_{j j'} (\alpha) = \dfrac{2^{\alpha}}{N}
    \sum_{k \in \mathcal{K}}
    e^{-2 \pi i (j-j') k/N}  \abs*{\sin^{\alpha} \left[ \tfrac{ \pi k}{N} \right]},
\end{equation}
with the DFT being defined as
\begin{equation} 
    \tilde{\phi}_{k} = \dfrac{1}{\sqrt{N}} \sum_{j \in \mathcal{J}} e^{- 2 \pi i j k/N} \phi_j,
    \label{eq:Discrete Fourier Transformation}
\end{equation}
and discrete momenta labeled by $k \in \mathcal{K} = \{0, \dots, N-1 \}$. Note here that discretization can become ambiguous for other than periodic boundary conditions~\cite{Lischke2020}.

\section{Quantum quench of the relativistic theory}

\subsection{Finite size effects}
\label{subsec:FiniteSizeEffects}

In a finite volume $L < \infty$, quasi-particles have discrete momenta and their dynamics are influenced by boundary conditions. We discuss the two most prominent choices: periodic and open boundaries.

\begin{figure*}[t!]
    \centering
    \includegraphics[width=0.99\textwidth]{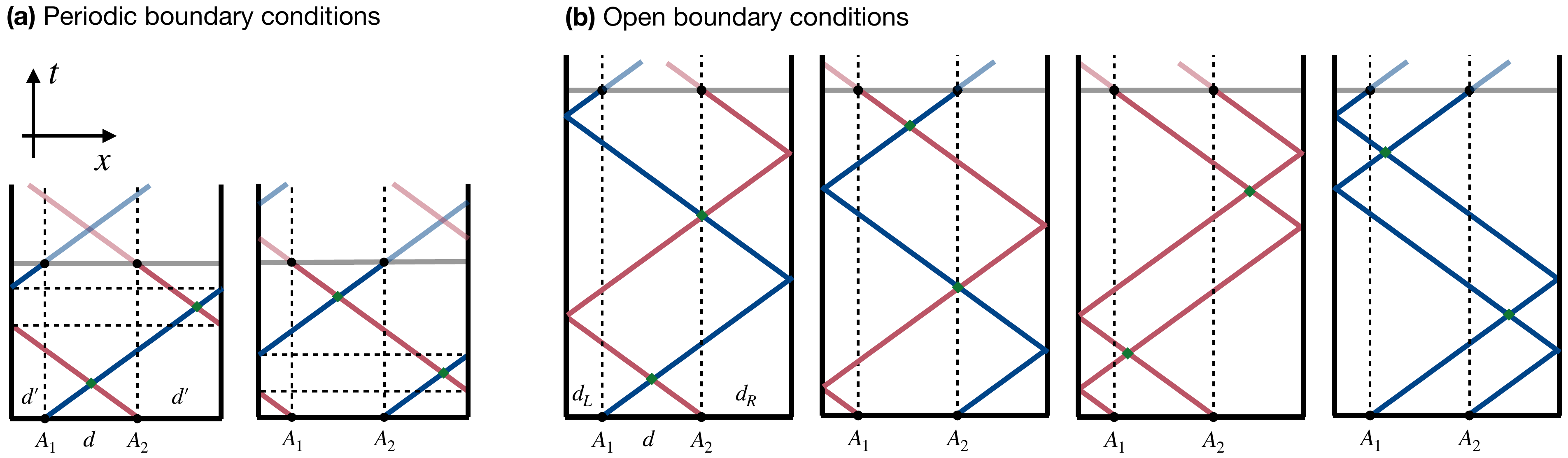}
    \caption{Boundary effects relevant for identifying the finite-size Rényi-2 mutual information. We trace all trajectories of two quasi-particles emerging at $A_1$ and $A_2$ that contribute to the buildup of correlations---marked by crossing trajectories.
    \textbf{(a)} Periodic boundary conditions. Parallel trajectories never cross. Hence, only particle pairs moving towards each other directly (left panel) or indirectly (right panel) contribute. This results in one (three) additional finite-size contribution(s) to the Rényi-2 entanglement entropy~\eqref{eq:Rényi-2 entropy in PBC} [mutual information~\eqref{eq:Rényi-2 MI in PBC}] compared to the infinite volume formulas presented in the main text. As the ballistic motion is periodic, we will observe momentum-dependent revivals.
    \textbf{(b)} Open boundary conditions. In addition to the previous two possibilities, quasi-particles moving in parallel now do contribute to correlations (two rightmost panels). Further, the periodicity doubles.}
    \label{fig:Pictorial mutual information in PBC and OBC}
\end{figure*}

\subsubsection{Periodic boundary conditions (PBCs)}
\label{subsubsec:PeriodicBoundaries}
For periodic boundaries, quasi-particles reenter from the left (right) when traveling to the right (left) and passing through the boundary at $L$ ($0$). The integral over momenta in~\eqref{eq:Renyi2EntropyQFT} is replaced by a sum and the condition $\min (2|v(p)|t, \ell)$, which can be expanded into the two terms $2|v(p)|t \times \theta (l - 2|v(p)|t) + \ell \times \Theta (2|v(p)|t - l)$, has to be modified to account for reentrances, see Fig.~\hyperref[fig:Pictorial mutual information in PBC and OBC]{9(a)}. Then, the Rényi-2 entanglement entropy becomes~\cite{Calabrese2020, Modak2020}
\begin{equation}
    \begin{split}
        S_2 (t, \ell) & = \sum_{g_{k} < \ell/L} s_{2,k} \, g_{k} + \dfrac{\ell}{L} \sum_{\ell/L \le g_{k} < 1 - \ell/L} s_{2,k} \\
        &\hspace{0.4cm} + \sum_{g_{k} \ge 1 - \ell/L} s_{2,k} (1 - g_{k}),
    \end{split}
    \label{eq:Rényi-2 entropy in PBC}
\end{equation}
where we sum over the momentum modes $k$ and introduced the function $g_{k} = \{ 2 |v_{k}| t /L \}$, with $\{ x \}$ denoting the fractional part of $x$. 

A similar argument can be made for adapting the Rényi-2 mutual information~\eqref{eq:Renyi2MI} to finite system sizes. Due to the circular geometry, the two subregions are separated by $d$ as well as the complementary distance $d' = L - 2 \ell - d$ as quasi-particles contributing to the mutual information can also move in opposite directions, yielding one additional contribution, see left vs. right panel in Fig.~\hyperref[fig:Pictorial mutual information in PBC and OBC]{9(a)}. Further, each quasi-particle pair contributes twice to the mutual information: once after having traveled a distance $(L + d)/2$, but also for $(L + d')/2$, see the green diamonds in Fig.~\hyperref[fig:Pictorial mutual information in PBC and OBC]{9(a)}. In total, there are \textit{four} contributions to mutual information before revivals, i.e., this process repeating itself. Hence, after introducing the function
\begin{equation}
    \begin{split}
    i_{2,k}(x) = s_{2,k} \Big[ &\max \left( x/L, g_k \right) \\
    &+ \max \left( x/L + 2 \ell/L, g_k \right) \\
    & - 2 \max \left(x/L + \ell/L, g_k \right) \Big],
    \end{split}
    \label{eq:i2Function}
\end{equation}
the finite-size mutual information takes the form
\begin{equation}
    \begin{split}
        I_{2}(t, \ell, d) = \sum_{k \in \mathcal{K}} \Big[& i_{2,k}(d) + i_{2,k}(d') \\[-2ex]
        & + i_{2,k}(L + d) + i_{2,k}(L + d') \Big].
        \label{eq:Rényi-2 MI in PBC}
    \end{split}
\end{equation}
Note that $\ell$ appears in the previous equation: Fig.~\ref{fig:Pictorial mutual information in PBC and OBC} sketches $M=1$, while we consider $|A_1|=|A_2|=M>1$.

\subsubsection{Open boundary conditions (OBCs)}
\label{subsubsec:OpenBoundaries}
For open boundaries, quasi-particles reaching the boundary are reflected, corresponding to a velocity inversion $v_k \rightarrow -v_k$. Whether the particles' phase jumps during elastic scattering or not is specified by Dirichlet (fixed ends $\phi_{0}=\phi_{N}=0$, phase jump of $\pi$) or von Neumann (free ends $(\phi_1 - \phi_0)/\epsilon = (\phi_N - \phi_{N-1})/\epsilon = 0$, no phase jump) boundary conditions. For our analysis, however, phase jumps are irrelevant. 

When considering the relativistic theory~\eqref{eq:Relativistic Real Lagrangian Density} and von Neumann boundaries, the  Hamiltonian matrix given in Eq.~\eqref{eq:Relativistic Real Hamiltonian Matrix Phi-Block} modifies to
\begin{equation} \label{eq:Relativistic Real Hamiltonian Matrix Phi-Block OBC}
    H_{\text{R}}^{\phi}
    = \begin{pmatrix}
        \mu + \nu & \nu &  \ldots & 0 & 0 \\
        \nu & \mu &  \ldots & 0 & 0 \\
        \vdots & \ddots & \ddots & \ddots & \vdots \\
        0 & 0 &  \ldots &\mu & \nu \\
        0 & 0 & \ldots & \nu & \mu + \nu
    \end{pmatrix} ,
\end{equation}
with $\mu = \epsilon m^{2} - 2 \nu$ and $\nu = - 1/\epsilon$ as before. The Hamiltonian is diagonalized by an orthogonal transformation resulting in the dispersion
\begin{equation} \label{eq:Dispersion relation real relativistic theory OBC}
    \omega^2_{k} = m^{2} + \dfrac{4}{\epsilon^{2}} \sin^{2} \left[ \tfrac{ \pi (k-1)}{2N} \right],
\end{equation}
which differs from the PBC result~\eqref{eq:Dispersion relation real relativistic theory} by a factor of $1/2$ in the argument of the sine. 

\begin{figure}[t]
    \centering
    \includegraphics[width=0.99\columnwidth]{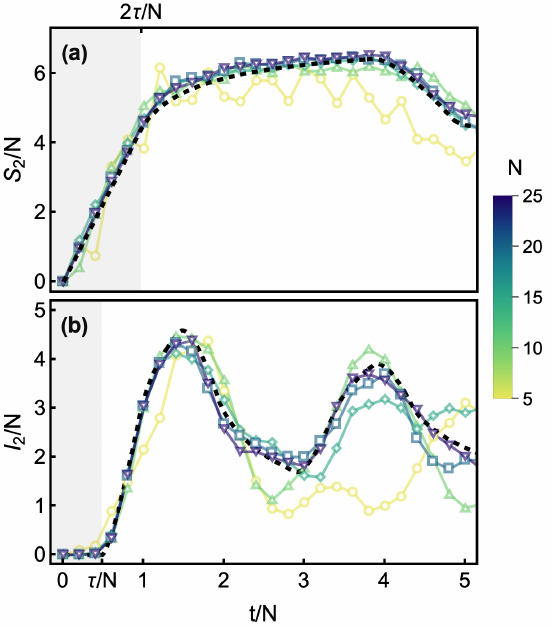}
    \caption{Same analysis as in Fig.~\ref{fig:EntanglementGeneration} for open boundaries (von Neumann) breaking circulanticity. \textbf{(a)} The Rényi-2 entanglement entropy builds up more slowly, with a linear increase until $2\tau/N = \ell/(Nv_{\text{max}}) \approx 0.98$ set by the fastest quasi-particles.
    \textbf{(b)} The Rényi-2 mutual information remains largely unaltered, especially for larger mode counts $N \gtrsim 20$.}
    \label{fig:EntanglementGenerationOBC}
\end{figure}

Also for open boundaries, the quasi-particle picture offers a reasonable description for how information spreads throughout the system. The discrete group velocity $v_k = \sin [\pi (k-1)/N]/(\epsilon \omega_k)$, although being of slightly different functional form, is almost identical to the PBC case for $N \gtrsim 10$. Hence, $v_{\text{max}}$ is unaltered. What changes are the quasi-particle trajectories, which impact the function $g_k$ entering~\eqref{eq:Rényi-2 entropy in PBC} and controlling the transition in the Rényi-2 entropy from linear growth (early times) to saturation (late times). For OBC, quasi-particles produced at distances $> \ell/2$ from the entangling region cannot enter the region $B$ by traversing the boundary, resulting in the replacement $g_k = \{ 2 |v_{k}| t /L \} \rightarrow \{ |v_{k}| t /L \}$, while~\eqref{eq:Rényi-2 entropy in PBC} remains intact. Therefore, the time until saturation doubles relative to PBC, i.e., $2 \tau/N$, which translates into a doubled revival time $L/v_{\text{max}}$, as shown in Fig.~\hyperref[fig:EntanglementGenerationOBC]{10(a)}.

For the Rényi-2 mutual information, we must account for multiple factors. Importantly, in this case, the function $g_k$ is the same one used in rthe PBC case. This follows after comparing the green diamonds in the left diagrams of Fig.~\hyperref[fig:Pictorial mutual information in PBC and OBC]{9(a)} and Fig.~\hyperref[fig:Pictorial mutual information in PBC and OBC]{9(b)}---the time required for the first build-up of correlations (and thus also the revival time) is identical. However, the number of possible trajectories entering the finite-size mutual information formula increases from four, as in Eq.~\eqref{eq:Rényi-2 MI in PBC}, to eight. Quasi-particles initially moving in parallel contribute to overall correlations after being reflected at the boundaries, see, e.g., the rightmost panel in Fig.~\hyperref[fig:Pictorial mutual information in PBC and OBC]{9(b)}. Moreover, the complementary distance $d'=L-2\ell-d$ considered for PBC, see left panel in Fig.~\hyperref[fig:Pictorial mutual information in PBC and OBC]{9(a)}, is not unique anymore. It must be replaced by $d_\text{L}$ and $d_\text{R}$, which denote the distances from the left region to the left wall and from the right region to the right wall, respectively, as shown in the leftmost panel of Fig.~\hyperref[fig:Pictorial mutual information in PBC and OBC]{9(b)}. For given $d_\text{L}$, one can compute $d_\text{R}$ via $d_\text{R} = L - 2\ell - d - d_\text{L}$. In summary, the finite-size mutual information for open boundaries attains the form
\begin{equation}
    \begin{split}
        &I_{2}(t, \ell, d, d_\text{L}) \\
        &= \sum_{k \in \mathcal{K}} \Big[ i_{2,k}(d) + i_{2,k}(L+d_\text{L}+d_\text{R}) \\[-2.2ex] 
        &\hspace{1.2cm} + i_{2,k}(2L + d) + i_{2,k}(3L+d_\text{L}+d_\text{R}) \\
        &\hspace{1.2cm}+ i_{2,k}(2d_\text{L} + d + \ell) + i_{2,k}(2d_\text{R} + d + \ell) \\
        &\hspace{1.2cm}+ i_{2,k}(2L+ 2d_\text{L} + d + \ell) \\[-1.1ex]
        &\hspace{1.2cm}+ i_{2,k}(2L+ 2d_\text{R} + d + \ell) \Big],
        \label{eq:Rényi-2 MI in OBC}
    \end{split}
\end{equation}
with $i_{2,k}(x)$ as in~\eqref{eq:i2Function}.

In Fig.~\ref{fig:EntanglementGenerationOBC}, we show the quench dynamics of the relativistic scalar as discussed in Sec.~\ref{subsec:Dynamical entanglement structure} with the same parameters as in Fig.~\ref{fig:EntanglementGeneration}, but open (von Neumann) instead of periodic boundaries. The early-time dynamics of the mutual information between two separated spatial regions essentially agree for periodic and open boundaries, i.e., $I_2^{\text{OBC}}(t,\ell,d) \simeq I_2^{\text{PBC}}(t,\ell,d)$  [compare Fig.~\hyperref[fig:EntanglementGenerationOBC]{10(b)} with Fig.~\hyperref[fig:EntanglementGeneration]{4(b)}], which we trace back to their similar first quasi-particle crossing times. Note, however, that their late-time evolution can differ. In contrast, the Rényi-2 entropy for OBC evolves at roughly half the pace of PBC, i.e., $S_2^{\text{OBC}}(t,\ell) \simeq S_2^{\text{PBC}}(t/2,\ell)$ [Fig.~\hyperref[fig:EntanglementGenerationOBC]{10(a)} vs. Fig.~\hyperref[fig:EntanglementGeneration]{4(a)}], since quasi-particles moving in opposite directions can no longer shortcut via periodic boundaries, which delays their crossings by a factor of two. As for our observations in Sec.~\ref{subsec:Dynamical entanglement structure}, the photonic simulations are in notable agreement with QFT predictions for similar system sizes $N \gtrsim 10-20$.

\subsection{Entropy density of the GGE and mode populations}
\label{subsec:Quantum quench population}
The entropy density $s_{2,k}$ follows from computing the Rényi-2 entanglement entropy~\eqref{eq:Renyi2Entropy} of the GGE $\boldsymbol{\rho}(\infty,l) \propto \text{exp}(-\sum_k y_k \boldsymbol{n}_k)$ in the eigenbasis of the quasi-particle number operator, leading to~\cite{Alba2018,Calabrese2020}
\begin{equation}
    S_2 (\infty, \ell) = S_{2,\text{GGE}} = \sum_{k \in \mathcal{K}} s_{2,k},
\end{equation}
with coefficients
\begin{equation}
    s_{2,k} = \ln \left[ - n_k^2 + ( 1 + n_k)^2 \right],
    \label{eq:Density entropy for bosonic system}
\end{equation}
where $n_k = \braket{0 | \boldsymbol{n}_k | 0}$ specifies the quasi-particle content of the initial state (that is conserved during time evolution). 

To compute $n_k$, we first determine the eigenmodes of the relativistic theory~\eqref{eq:Relativistic Real Hamiltonian Operator in Real Space}. As discussed in Sec.~\ref{subsec:Relativistic Real Field}, performing a DFT~\eqref{eq:Discrete Fourier Transformation} to momentum space renders the corresponding Hamiltonian diagonal, to wit
\begin{equation}
    \boldsymbol{H}_{\text{R}} = \sum_{k \in \mathcal{K}} \omega_{k} \left(\tilde{\boldsymbol{a}}_{k}^{\dagger} \tilde{\boldsymbol{a}}_{k} + \dfrac{1}{2} \right). 
    \label{eq:RelativisticHamiltonianDiagonal}
\end{equation}
Therein, we introduced the mode operators
\begin{equation}
    \begin{split}
    \tilde{\boldsymbol{a}}_{k} &= \dfrac{1}{\sqrt{2 \epsilon \omega_{k}}}\left( \epsilon \omega_{k} \tilde{\boldsymbol{\phi}}_{k} + i \tilde{\boldsymbol{\pi}}_{k} \right), \\
    \tilde{\boldsymbol{a}}_{k}^{\dagger} &= \dfrac{1}{\sqrt{2 \epsilon \omega_{k}}} \left( \epsilon \omega_{k} \tilde{\boldsymbol{\phi}}_{k}^{\dagger} - i \tilde{\boldsymbol{\pi}}_{k}^{\dagger} \right) ,
    \end{split}
    \label{eq:ModeOperatorsQuasiParticles}
\end{equation}
describing quasi-particle excitations. Note here that the appearance of the relativistic dispersion $\omega_k$ guarantees Lorentz-invariance of the momentum integral measure in the continuum and necessitates squeezing.

The ground state $\ket{\Omega}$ of~\eqref{eq:Relativistic Real Hamiltonian Operator in Real Space}, or, equivalently, \eqref{eq:RelativisticHamiltonianDiagonal}, corresponds to the quasi-particle vacuum $\tilde{\boldsymbol{a}}_k \ket{\Omega} = 0$, while the input-state vacuum $\boldsymbol{a}_j \ket{0} = 0$ contains $n_k$ quasi-particles. The latter follows after expanding the quasi-particle mode operators $(\tilde{\boldsymbol{a}}_{k}, \tilde{\boldsymbol{a}}_{k}^{\dagger})$ in terms of the position-space, i.e., photonic, mode operators $(\boldsymbol{a}_j,\boldsymbol{a}_j^{\dagger})$, for which we combine Eqs.~\eqref{eq:ModeOperatorsQuasiParticles} and~\eqref{eq:Discrete Fourier Transformation} with $\boldsymbol{a}_j = (\boldsymbol{\phi}_j + i \boldsymbol{\pi}_j)/\sqrt{2}$, leading to
\begin{equation}
    \tilde{\boldsymbol{a}}_{k} = \dfrac{1}{2 \sqrt{N}} \sum_{j \in \mathcal{J}} e^{-2\pi i j k/N}
    ( \eta_{k} \boldsymbol{a}_{j}
    + \kappa_{k} \boldsymbol{a}_{j}^{\dagger} ) ,
\end{equation}
with
\begin{equation}
    \eta_{k} = \sqrt{\epsilon \omega_{k}} + \dfrac{1}{\sqrt{\epsilon \omega_{k}}}, \quad \kappa_{k} = \sqrt{\epsilon \omega_{k}} - \dfrac{1}{\sqrt{\epsilon \omega_{k}}} .
\end{equation}
This way, we find
\begin{equation} \label{eq: Number of quasi-particles}
    \begin{split}
        n_{k}
        & = \braket{ 0 | \tilde{\boldsymbol{a}}_{k}^{\dagger} \tilde{\boldsymbol{a}}_{k} | 0} \\
        & = \dfrac{\kappa_{k}^{2}}{4 N}
        \sum_{j,j' \in \mathcal{J}}
        e^{-2 \pi i (j - j') k/N} 
        \langle 0 | \boldsymbol{a}_{j'} \boldsymbol{a}_{j}^{\dagger} | 0 \rangle \\
        &= \dfrac{\kappa_{k}^{2}}{4} \\
        &= \dfrac{1}{4} \left( \epsilon \omega_{k} + \dfrac{1}{\epsilon \omega_{k}} - 2 \right) .
    \end{split}
\end{equation}

\subsection{Comment on divergences in the continuum limit}
\label{subsec:Divergences}
The input state chosen in the main text, which is experimentally motivated, leads to divergences in the continuum. Formally, the corresponding pre-quench Hamiltonian~\eqref{eq:PreQuenchHamiltonian} attains the form
\begin{equation}
    \boldsymbol{H}_{0} = \dfrac{1}{2} \int \mathrm{d}x \left[ \boldsymbol{\pi}^{2} + (\lim_{\epsilon \rightarrow 0} \epsilon^{-2})
     \boldsymbol{\phi}^{2} \right]
\end{equation}
in the continuum. It becomes apparent that this Hamiltonian has a divergent mass term. This causes further divergences in the corresponding momentum modes' energy ($\omega_{0,k} = 1/\epsilon$), population numbers ($n_k \sim 1/\epsilon$), and thus also the entropy density $s_{2,k} \sim - \ln \epsilon$. We remark that, as of working within a regularized theory, these divergences can be cured by renormalizing the considered quantities with appropriate functions of $\epsilon$.


\bibliography{references.bib}

\end{document}